\documentclass{emulateapj}

\usepackage{graphicx}
\usepackage{times}
\usepackage{natbib}
\usepackage{url}
\usepackage{color}
\usepackage{amssymb,amsmath}

\usepackage{graphicx}
\usepackage{dcolumn}
\usepackage{bm}

\def\lapprox{\mathrel{\hbox{\rlap{\hbox{\lower4pt\hbox{$\sim$}}}\hbox{$<$}}}}
\def\gapprox{\mathrel{\hbox{\rlap{\hbox{\lower4pt\hbox{$\sim$}}}\hbox{$>$}}}}

\newcommand{\be}{\begin{equation}}
\newcommand{\ee}{\end{equation}}

\newcommand {\nind} {\noindent}
\newcommand {\mb} {\mathbf}

\newcommand {\bea} {\begin{eqnarray}}
\newcommand {\eea} {\end{eqnarray}}

\begin{document}

\title{Simulations of Emerging Magnetic Flux. I: The Formation of Stable Coronal Flux Ropes}

\author{James E. Leake}
\email{jleake@gmu.edu}
\affiliation{College of Science, George Mason University, 4400 University Drive, Fairfax, Virginia 22030.}
\author{Mark G. Linton}%
\affiliation{ U.S. Naval Research Laboratory 4555 Overlook Ave., SW Washington, DC 20375.}
\author{Tibor T\"{o}r\"{o}k}%
\affiliation{Predictive Science Inc., 9990 Mesa Rim Rd., Ste. 170, San Diego, CA 92121.}

\date{\today}
\begin{abstract}

We present results from 3D visco-resistive magnetohydrodynamic (MHD) simulations of the emergence of a convection zone magnetic flux tube into a solar atmosphere containing a pre-existing dipole coronal field, which is orientated to minimize reconnection with the emerging field. We observe that the emergence process is capable of producing a coronal flux rope by the transfer of twist from the convection zone as found in previous simulations. We find that this flux rope is stable, with no evidence of a fast rise, and that its ultimate height in the corona is determined by the strength of the pre-existing dipole field. We also find that although the electric currents  in the initial convection zone flux tube are almost perfectly neutralized, the resultant coronal flux rope carries a significant net current. These results suggest that flux tube emergence is capable of creating non-current-neutralized stable flux ropes in the corona, tethered by overlying potential fields, a magnetic configuration that is believed to be the source of coronal mass ejections.

\end{abstract}

\maketitle

\section{Introduction}

Coronal mass ejections (CMEs) are a primary source of space weather and almost all theoretical models of CMEs require the presence or formation of a coronal magnetic flux rope \citep[e.g.,][]{2000JGR...10523153F}. There exists observational evidence that many CMEs, particularly those originating from quiet Sun regions, are composed of a bright core associated with an erupting prominence, and a relatively dark cavity that is associated with a magnetic flux rope \citep[e.g.,][]{2006JGRA..11112103G}. Moreover, a magnetic flux rope geometry has been fitted to coronagraph observations of propagating CMEs \citep[e.g.,][]{2012SoPh..tmp..192V}. In addition, there is also growing evidence that these flux ropes are formed \textit{before} the eruption. This evidence exists for both quiet Sun regions \citep{Robbrecht_2009, Vourlidas_Lynch_2012}, and active regions \citep[e.g.,][]{Green_2009,Green_2011,Patsourakos_2013}. Although identification of flux rope magnetic geometry is more difficult for active regions than for the quiet Sun due to differences in size and complexity, the existence of active region flux ropes  has been supported by recent non-linear force free extrapolations \citep[e.g.,][]{Canou_2010,Guo_2010,2012ApJ...748...23Y}.

It has long been postulated that the source of the magnetic field in the corona is magnetic field in the deep convection zone, created by dynamo action \citep{1979ApJ...230..905P}, and that the process by which this field arrives in the corona is the buoyant rise of twisted flux tubes to the surface, and their subsequent emergence. 
The partial emergence of twisted flux tubes into the solar atmosphere has been extensively studied, and a review by \citet{2008JGRA..11303S04A} summarizes the various types of theoretical investigations. Early 3D simulations found that an emerging sub-surface flux tube does not rise bodily into the corona, but that only the upper portion of the tube emerges, while the tube axis remains near the solar surface \citep{2001ApJ...554L.111F,2001ApJ...549..608M,2004A&A...426.1047A,2006A&A...460..909M}. More recent simulations found that a new flux rope structure forms in the corona within the partially emerged flux tube, and this flux rope rises slowly in the corona \citep{2004ApJ...610..588M,Fan_2009}. \citet{2008A&A...492L..35A} and \citet{Archontis_Hood_2012}  demonstrated that the rise of the  flux rope came to a halt, due to the stabilizing magnetic tension of the surrounding (envelope) flux tube field. \citet{Fan_2009} associated the coronal flux rope formation mechanism with the transfer of twist from the convection zone, while \citet{2004ApJ...610..588M} suggested that the mechanism is due to the reconnection of sheared magnetic fields. By imposing a pre-existing strong horizontal field in the corona, \citet{2006ApJ...645L.161A} found that reconnection between the emerging flux tube and the coronal field can create horizontal jets and plasmoids at relatively low heights in the corona. Later simulations also found that favorably orientated and sufficiently weak horizontal fields can remove part of the envelope field constraining the newly formed flux rope, allowing a strong upward acceleration of the rope resembling eruptive behavior \citep{2008A&A...492L..35A,Archontis_Hood_2012}. 

In this paper we focus  on the formation of stable coronal flux ropes as a result of flux emergence. Creating such configurations is  an important step in improving initial equilibrium magnetic field configurations for models of CMEs. A primary example of a pre-eruption configuration is the flux rope model of \citet{1999A&A...351..707T} (hereafter the TD model) in which a coronal flux rope is confined by an overlying potential field. This model has been successfully applied as the initial condition of a number of CME simulations \citep{2003ApJ...588L..45R,2005ApJ...630L..97T,2008ApJ...684.1448M}. A specific property of the TD  model is that the coronal flux rope carries a net current, since there is no return current in the configuration (see \citet{Torok_2013} for a detailed discussion of return currents in active regions).

Our aim in this paper is to use numerical magnetohydrodynamic (MHD) simulations to investigate how flux emergence from the convection zone into the solar atmosphere can create a stable coronal flux rope with a net current that is confined by overlying magnetic field. To do this we model the partial emergence of a buoyant convection zone twisted flux tube into a pre-existing dipole coronal magnetic field, the strength of which we vary. The coronal dipole field is intended to represent the remnant field of an old, dispersed active region, into which new magnetic field emerges. This simulation of flux emergence into a pre-existing dipole follows a paradigm similar to that of the simulation presented in \citet{Mactaggart_2011}. That study focused on reconnection between emerging field and pre-existing field in the corona, whereas in this study we focus on the formation and stability of a coronal flux rope formed within the emerging field. In addition, for reference, we also model flux emergence into a field-free corona, to compare to the simulations of   \citet{Archontis_Hood_2012} and \citet{2004ApJ...610..588M}.

In Section \ref{Num} the model is described. The results are presented in Section \ref{Results}, and the consequences of these simulations for the theory of coronal flux rope formation and CME initiation are discussed in Section \ref{Disc}.

\section{Numerical Method}
\label{Num}
\subsection{Equations}

\begin{figure}[t]
\centering
\includegraphics[width=1.\linewidth]{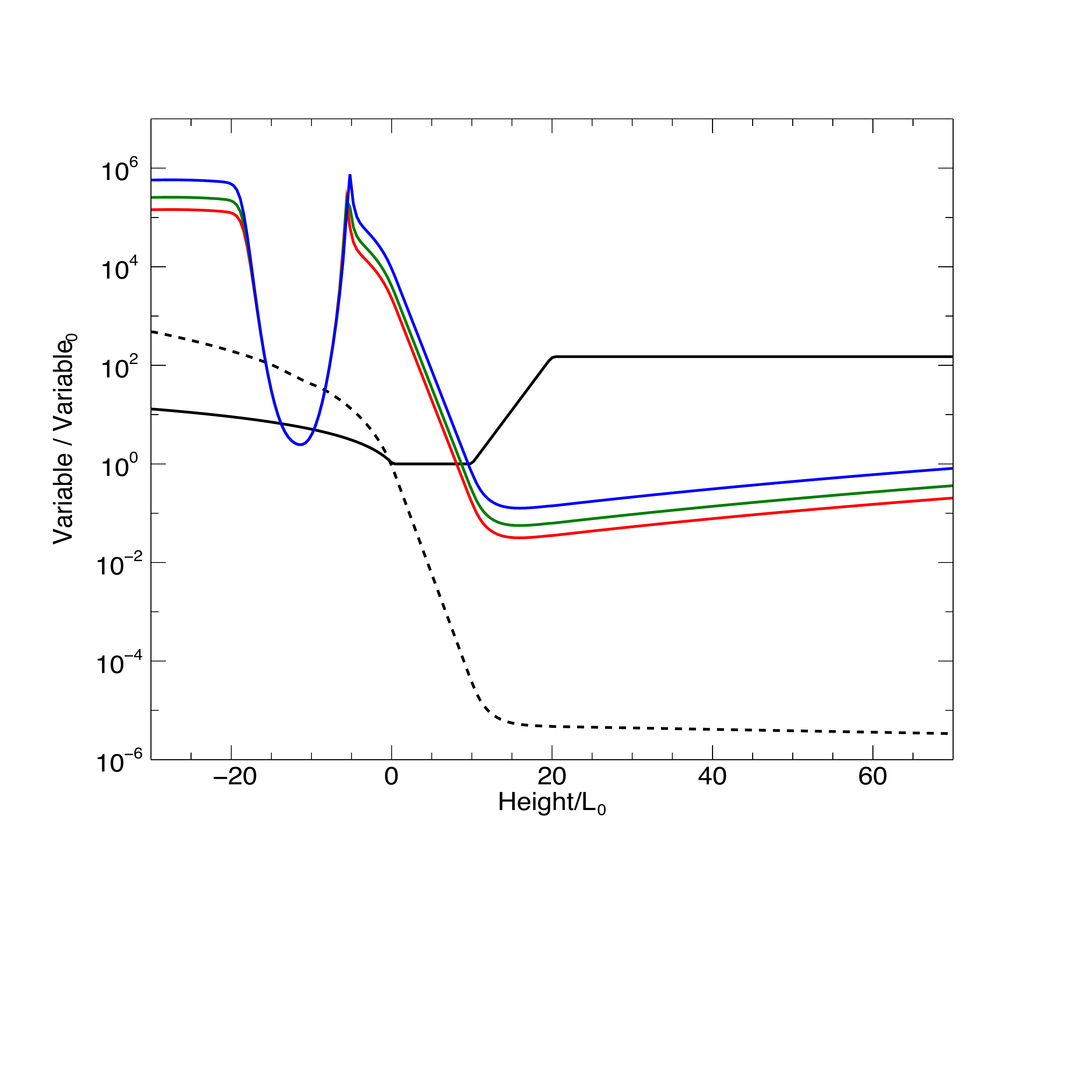}
\vspace{-25mm}
\caption{Initial 1D configuration (along $x=y=0$) for part of the vertical domain. The ordinate represents normalized plasma variables ($C/C_{0}$ for a given variable $C$), and the abscissa represents height. The black solid line shows the temperature. The black dashed line shows gas pressure $P$. The red, green, and blue lines show the plasma-$\beta$ for the three simulations SD, MD, and WD, respectively. These three simulations have decreasing dipole field strengths.
\label{fig:IC_1D}}
\end{figure}

The evolution of magnetic field in a plasma domain which includes the upper layers of the solar convection zone, plus a photosphere/chromosphere, transition region and corona, is modeled using the visco-resistive magnetohydrodynamics (MHD) Lagrangian-remap code Lare3D \citep{2001JCoPh.171..151A}.  The equations solved by Lare3D are presented here in Lagrangian form:

\begin{eqnarray}
\frac{D\rho}{Dt} & = & -\rho\nabla\cdot\mathbf{v}, \\
\frac{D\mathbf{v}}{Dt} & = & -\frac{1}{\rho}\left[\nabla P 
+ \mathbf{j}\times\mathbf{B} + \rho\mathbf{g} + \nabla\cdot\mathcal{S}\right],\label{eqn:vel}\\
\frac{D\mathbf{B}}{Dt} & = & (\mathbf{B}\cdot\nabla)\mathbf{v} 
- \mathbf{B}(\nabla \cdot\mathbf{v}) - \nabla \wedge (\eta\mathbf{j}), \textrm{and} \\
\frac{D\epsilon}{Dt} & = & \frac{1}{\rho}\left[-P\nabla \cdot\mathbf{v}
 + \varsigma_{ij}\mathcal{S}_{ij} + \eta {j}^{2}\right].
\label{eqn:energy_MHD}
\end{eqnarray}
Here $\rho$ is the mass density, $\mathbf{v}$ the velocity, $\mathbf{B}$ the magnetic field, and $\epsilon$ the specific energy density. The current density is given by $\mathbf{j}=\nabla\times\mathbf{B}/\mu_{0}$, $\mu_{0}$ is the permeability of free space, and the resistivity $\eta=14.6 ~ \Omega~\textrm{m}$. The gravitational acceleration is denoted by $\mathbf{g}$ and is set to the gravity at the mean solar surface ($\mb{g}_{sun} = -274 ~ \textrm{m}\textrm~{s}^{-2}\hat{\mb{z}}$). $\mathcal{S}$ is the stress tensor which has components 
$\mathcal{S}_{ij}=\nu(\varsigma_{ij}-\frac{1}{3}\delta_{ij}\nabla\cdot\mathbf{v})$, with
$\varsigma_{ij}=\frac{1}{2}(\frac{\partial v_{i}}{\partial x_{j}}+
\frac{\partial v_{j}}{\partial x_{i}}).$ The viscosity $\nu$ is set to $3.35\times10^{3} ~ \textrm{kg}~\textrm{m}^{-1}\textrm{s}^{-1}$, and $\delta_{ij}$ is the Kronecker delta function. Assuming an ideal gas law, the gas pressure, $P$, and the specific internal energy density, $\epsilon$, are
\begin{eqnarray}
P & = & \rho k_{B}T/\mu_{m}, ~ \mathrm{and} \\
\epsilon & =  & \frac{k_{B}T}{\mu_{m}(\gamma-1)},
\label{eqn:eos}
\end{eqnarray}
respectively, where $k_{B}$ is Boltzmann's constant and  $\gamma$ is 5/3. The reduced mass, $\mu_{m}$, is given by $\mu_{m}=m_{f}m_{p}$, where $m_{p}$ is the mass of a proton and $m_{f}=1.25$ is a pre-factor designed to include the effect of  elements heavier  than hydrogen.  In this study, as in many previous simulations of flux emergence, we assume that the plasma is fully ionized and so the reduced mass is spatially independent. In the partially ionized plasma of the Sun, the reduced mass changes as the ionization changes, as discussed in 
 \citet{2013ApJ...764...54L}. In this study however, we use the average value of $\mu_{m}=m_{f}m_{p}$, which was shown in \citet{2013ApJ...764...54L} to give the best constant-$\mu_{m}$ match to 1D models of the solar atmosphere that include partial ionization effects \citep[e.g.,][]{1981ApJS...45..635V, 2006ApJ...639..441F}. 
The plasma variables $\epsilon$ and $\rho$ are defined at cell centers. The magnetic field is defined at cell faces, and the velocity is defined at cell vertices. The staggered grid preserves 
$\nabla\cdot\mathbf{B}$ during the simulation. 

\begin{figure*}[t]
\centering
\includegraphics[width=1. \linewidth]{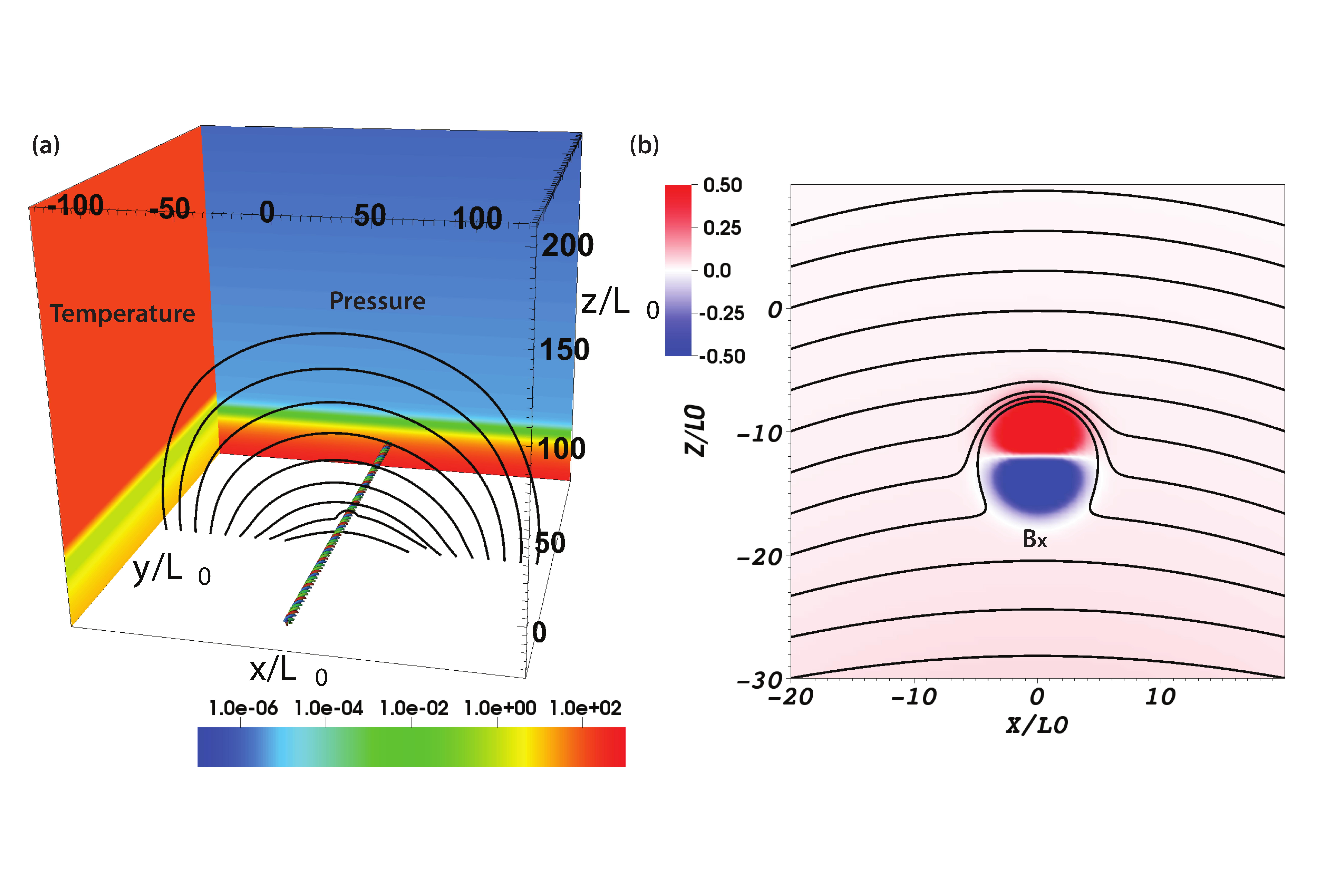}
\vspace{-20mm}
\caption{Initial 3D configuration for Simulation SD. Panel (a) shows the initial sub-surface flux tube and the dipole field, represented by the black fieldlines which originate from a line along $y=0$ on the bottom boundary. Panel (b) shows a magnified slice of the horizontal field $B_{x}$ in the $y=0$ plane as a color shading, and a projection onto the $y=0$ plane of the fieldlines of the dipole. \label{fig:IC_3D}}
\end{figure*}

\subsection{Normalization}

The equations are non-dimensionalized 
by dividing each variable ($C$) by its normalizing value ($C_{0}$).
The set of equations requires a choice of three normalizing values. We choose normalizing values for the length ($L_{0}=1.7\times10^{5} ~ \textrm{m}$),  magnetic field ($B_{0}=0.12 ~ \textrm{T} $), and gravitational acceleration ($g_{0}=g_{sun}=274  ~ \textrm{m}~\textrm{s}^{-2}$).  From these values the normalizing values for the gas pressure  ($P_{0}=B_{0}^2/\mu_{0}= 1.14	\times10^{4} ~ \textrm{Pa}$), density ($\rho_{0}=B_{0}^2/(\mu_{0}L_{0}g_{0})=2.46\times10^{-4} ~ \textrm{kg}~\textrm{m}^{-3}$), velocity ($v_{0}=\sqrt{L_{0}g_{0}}=6.82\times10^{3} ~ \textrm{m}~\textrm{s}^{-1}$), time ($t_{0}=\sqrt{L_{0}/g_{0}}$ =24.9 s), temperature ($T_{0}=m_{p}L_{0}g_{0}/k_{B} = 5.64\times10^{3}~\textrm{K}$), 
current density ($j_{0}=B_{0}/(\mu_{0}L_{0})=0.56 ~ \textrm{A}~\textrm{m}^{2}$), 
viscosity ($\nu_{0} = B_{0}^{2}\sqrt{L_{0}/g_{0}}/\mu_{0}=2.85\times10^{5} ~ \textrm{kg}~\textrm{m}^{-2}\textrm{s}^{-1}$), and resistivity ($\eta_{0} = \mu_{0}L_{0}^{\frac{3}{2}}g_{0}^{\frac{1}{2}} = 1.46\times10^{3} ~ \Omega~$m) can be derived. With these values of normalization, and the values of $\nu$ and $\eta$ given above,  the Reynolds number  $R_{e}=(\rho_{0}L_{0}v_{0})/\nu$ and magnetic Reynolds number $R_{m} = (\mu_{0}L_{0}v_{0})/\eta$ in this simulation are both 100. The value of resistivity used in this simulation ($0.01\eta_{0}=14.6 ~ \Omega ~\textrm{m}$), although comparable to upper estimates of the resistivity in the lower chromosphere, is much larger than typical values in the corona. 
We use this large value to ensure that in regions where electric current densities build up, the explicit resistivity is larger than the numerical value for the scheme used. The normalized numerical resistivity is $\hat{v}_{A}{\hat{\Delta}_{x}}^{2}/\hat{L}$ where $\hat{\Delta}_{x}$ is the normalized grid size, $\hat{v}_{A}$ is the normalized local Alfven speed, and $\hat{L}$ is a typical normalized length scale over which the magnetic field varies \citep{2007ApJ...666..541A}. In regions of increased current density, we find $\hat{\Delta}_{x}= 0.66$, $\hat{v}_{A}=0.05$, and $\hat{L}=5$, which gives a value for the normalized numerical resistivity of $\eta/\eta_{0} = 0.0044$.

\begin{figure*}[t]
\centering
\includegraphics[width=\linewidth]{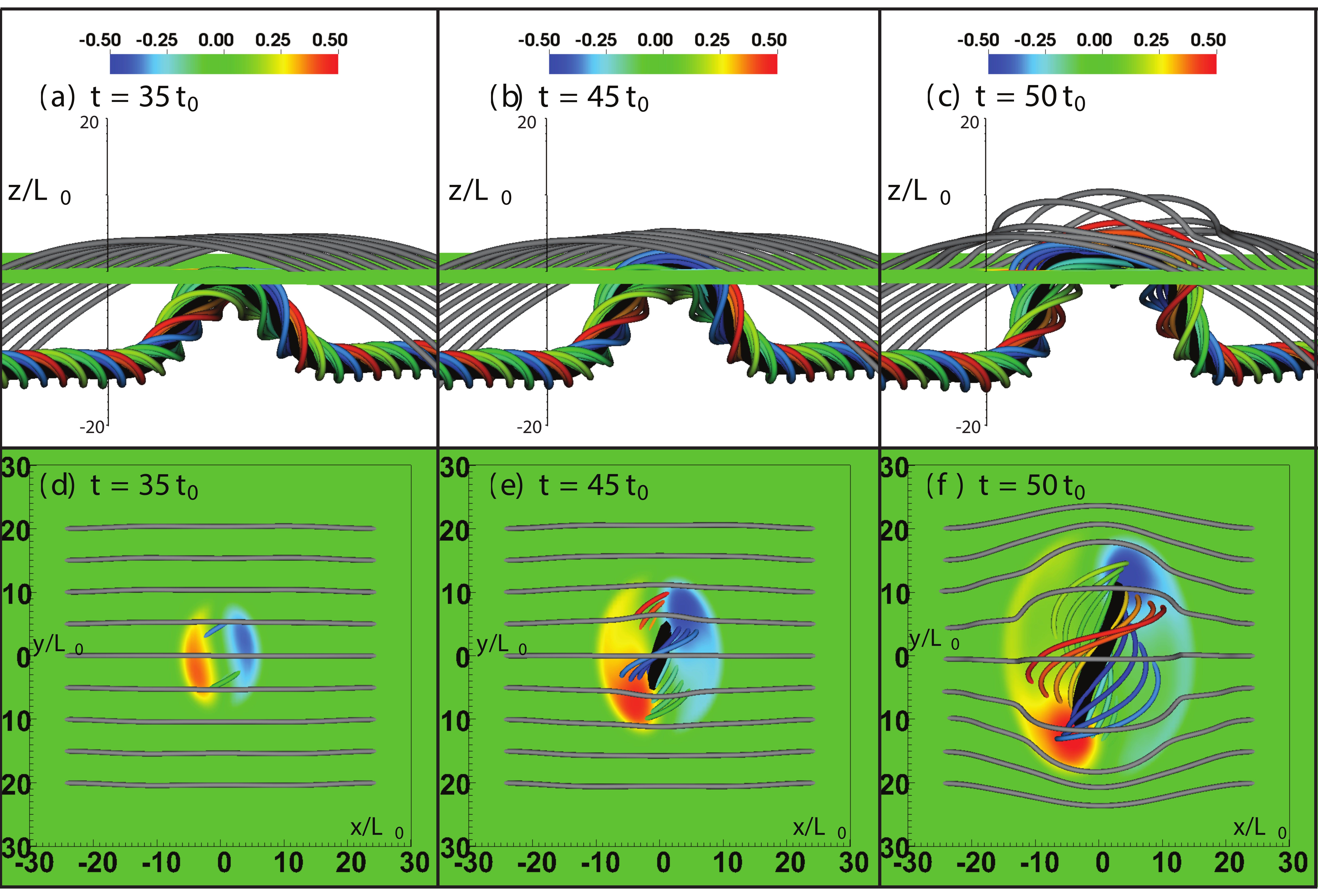}
\vspace{-0mm}
\caption{Early emergence of a convection zone magnetic flux tube into the solar atmosphere for Simulation SD at times $35t_{0}$ (Panels (a) and (d)), $45t_{0}$ (Panels (b) and (e)), and $50t_{0}$ (Panels (c) and (f)). 
The black line originates from the location of the original flux tube axis on the side ($y=\pm\max{y}$) boundaries. The colored lines originate from a circle on both the side boundaries centered on this axis. The grey lines originate at the lower boundary and belong to the dipole field. The color shading slice shows vertical magnetic field $B_{z}$ at the surface ($z=0$). This figure illustrates that the apparent shearing of the bipolar structure is associated with the emergence of the flux tube's axis and the magnetic field which is more aligned with this axis.   \label{fig:Early_FE_1}}
\end{figure*}

\subsection{Numerical Domain}

The simulations use an irregular cartesian grid with 304 cells in each direction. In the vertical direction, $z$, the grid extends from $-30L_{0}$ to $210.45L_{0}$ with a resolution of $0.428L_{0}$ at the bottom boundary and $1.99L_{0}$ at the top boundary. In the horizontal directions, $x$ and $y$, the grid is centered on $0$ and has side boundaries at $\pm 126.85L_{0}$. The resolution at $x=y=0$ is $0.658L_{0}$, and at the side boundaries is $2.61L_{0}$. This irregular grid has the following form:
 \begin{eqnarray}
x,~y = \pm \left[(1+f_{h})\chi_h + f_{h}w\ln\left(\frac{\cosh(\frac{\chi_h-L_{h}}{w})}{\cosh(\frac{-L_{h}}{w})}\right)\right] \\
z   =  -30L_{0} + \left[(1+f_{v})\chi_v + f_{v}w\ln\left(\frac{\cosh(\frac{\chi_v-L_{v}}{w})}{\cosh(\frac{-L_{v}}{w})}\right)\right]
\end{eqnarray}
where $\chi_h=[0,1,2...,152]100L_{0}/152$,  $\chi_v=[0,1,2,...,304]130L_{0}/304$, $f_{h}=2.1$, $f_{v}=1.83$, $L_{h}=95L_{0}$, $L_{v}=100L_{0}$, and $w=10L_{0}$. We also perform one additional simulation (named ND1) which has a higher top boundary at $270L_{0}$, with $f_{v}=2.83$.

At the boundaries all components of the velocity, and the gradients of magnetic field, gas density, 
and specific energy density are set to zero. The resistivity is smoothly decreased to zero close to the side boundary to reduce diffusion of magnetic field at the boundary to its numerical value and ensure line-tied boundary conditions as much as is possible:
\be
\eta = 0.01\left[\tanh(-\frac{(r_{\eta}-L_{\eta})}{w_{\eta}})+1\right]\frac{\eta_{0}}{2},
\ee
where $r_{\eta} = \sqrt{x^2+y^2}$, $L_{\eta}=100L_{0}$ and $w_{\eta}=5L_{0}$. In addition, a damping region is applied to the velocity at all four side boundaries and the top boundary. For a given coordinate ($\kappa=x,y,z$) the velocity equation (\ref{eqn:vel}) has an additional term when $|\kappa|>\kappa_{d}$:
\be
\frac{D\mathbf{v}}{Dt} = -\frac{1}{\rho}\left[\nabla P 
+ \mathbf{j}\wedge\mathbf{B} + \rho\mathbf{g} + \nabla\cdot\mathcal{S}\right] - N\mathbf{v},
\ee
with $x_{d}=y_{d}=96L_{0}$ and $z_{d}=170L_{0}$ (Simulation ND1 has $z_{d1}=254L_{0}$). The parameter $N$ is designed to increase linearly from 0 at $\kappa_{d}$ to 1 at the boundary: $N=(|\kappa|-\kappa_{d})/(\max{|\kappa|}- \kappa_{d})$.  This approach is used to prevent any reflected waves from interfering with the solution in the interior.

\subsection{Initial Conditions}

The initial conditions consist of a hydrostatic background atmosphere that represents the upper solar convection zone ($-30L_{0}\le z< 0$), the photosphere/chromosphere ($0\le z< 10L_{0}$), the transition region ($10L_{0}\le z < 20L_{0}$), and the corona ($20L_{0} \le z$). { The transition region in this model is thicker than a typical width derived from semi-implicit models of the Sun, which compare the observed spectrum of the Sun with radiative transfer calculations \cite[e.g.][]{2006ApJ...639..441F}. In those studies, the typical width is about 0.1 Mm ($ \approx L_{0}$). As in previous simulations of flux emergence \cite[see the review by][]{2008JGRA..11303S04A}, we use a thicker transition region of ~ 1.7 Mm.  This artificial increase of the transition region is required to resolve the large changes in density and temperature that occur across this region for the given spatial resolution which is limited by the large total simulation domain and computational costs.} A magnetic field is imposed on this background atmosphere. This field consists of a background dipole field that permeates the entire domain, and a localized twisted flux tube in the model convection zone. The flux tube's pressure and density are perturbed to initiate its buoyant rise into the model solar atmosphere. The initial conditions are shown in Figures \ref{fig:IC_1D} and \ref{fig:IC_3D}. 

\begin{figure*}[t]
\centering
\includegraphics[width=\linewidth]{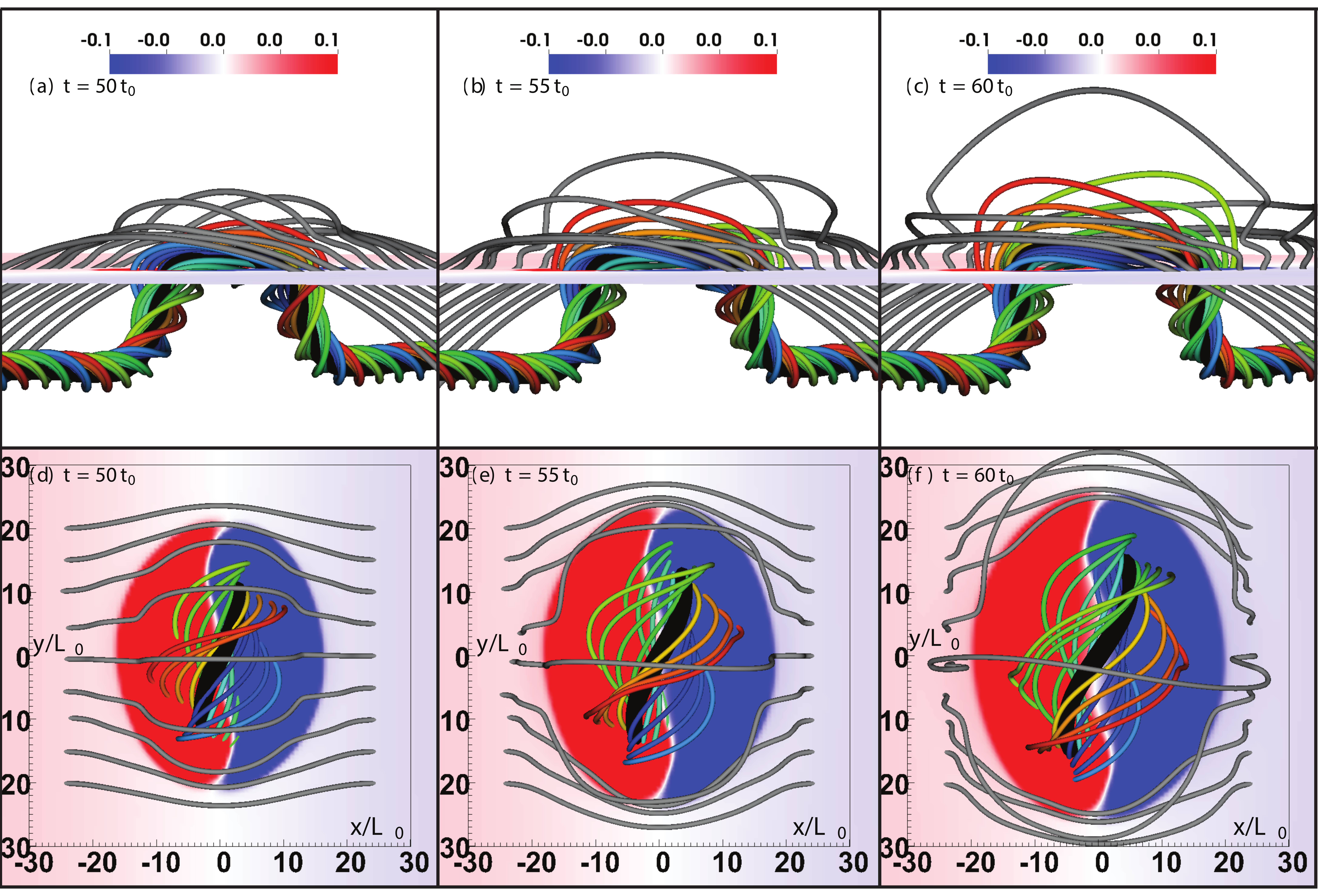}
\vspace{-0mm}
\caption{Partial emergence of the convection zone flux tube for Simulation SD at times $50t_{0}$ (Panels (a) and (d)), $55t_{0}$ (Panels (b) and (e)), and $60t_{0}$ (Panels (c) and (f)). The fieldlines shown are the same as in Figure \ref{fig:Early_FE_1}. The color shading of vertical magnetic field at the surface has been saturated to highlight the neutral line (which is located in white regions between red and blue regions). 
Above the flux tube axis (black line), the fieldlines are concave-down, and are able to drain mass and continue to rise. Below the axis, the fieldlines are concave up, and are unable to drain mass, and hence remain near the surface. Consequentially, only the upper part of the flux tube emerges into the corona. \label{fig:Early_FE_2}}
\end{figure*}

The initial hydrostatic atmosphere is created by first defining the temperature:
\bea
\frac{dT}{dz} & = & a\left(\frac{dT}{dz}\right)_{ad} = -\frac{\gamma-1}{\gamma}\frac{T_{0}}{L_{0}}, ~ z \le 0 ~ ; \label{eqn:dtdz}\\
T(z) & = & T_{ph}, ~  0 < z < 10L_{0} ~ ; \\
T(z) & = & T_{cor}^{(z-10L_{0})/10L_{0}},~ 10L_{0} \le z< 20L_{0} ~ ; \\
T(z) & = & T_{cor}, ~ z \ge 20L_{0} ~ ; 
\eea
where $T_{ph}=T_{0}$, $T_{cor}=150T_{0}$.
The pre-factor $a=1$ in Equation (\ref{eqn:dtdz}) ensures that the model convection zone is marginally stable to convective instability by setting the temperature gradient  to its adiabatic value $\frac{dT}{dz}=\left(\frac{dT}{dz}\right)_{ad}$ \citep{STIX}. The gas density profile is then obtained by solving the hydrostatic equilibrium equation $\partial P /\partial z = -\rho g$ and using the ideal gas law and the condition that $\rho(z=0) = \rho_{0}$.

The dipole field is translationally invariant along $y$, the tube's axial direction, and is given by $\mb{B}=\nabla \times \mb{A}$ where $\mathbf{A} = A_{y}\mathbf{e}_{y}$ and
\be
A_{y}(x,z) = B_{d}\frac{z-z_{d}}{r_{1}^{3}},
\ee
with $r_{1}=\sqrt{x^2+(z-z_{d})^2}$ being the distance from the source. We choose $z_{d}$ to be $-100L_{0}$ so that the initial sub-surface flux tube is far from the source of the 
dipole field. To cover various dipole strengths, we perform three simulations each with a different value of $B_{d}$. Simulations SD (strong dipole), MD (medium dipole) and WD (weak dipole) have values $B_{d} = [10,7.5,5]\times10^{3}B_{d0}$ where $B_{d0}=B_{0}{L_{0}}^2=3.76\times10^{9} ~ \textrm{T}~\textrm{m}^{2}$, respectively. This gives a maximum magnetic field strength at the surface ($z=0$) of $[2.6, 1.95, 1.3]\times10^{-3}$ T, respectively. These choices of dipole strength allow for a range in the plasma-$\beta$ profile, as shown in Figure \ref{fig:IC_1D}, where $\beta=\frac{P}{|\mb{B}|^{2}/\mu_{0}}$. These profiles are consistent with the models of $\beta$ in the solar atmosphere developed by \citet{1999SoPh..186..123G} and \citet{ 2001SoPh..203...71G}. Simulations ND and ND1 have no pre-existing dipole field.

\begin{figure*}[t]
\centering
\includegraphics[width=0.85\linewidth]{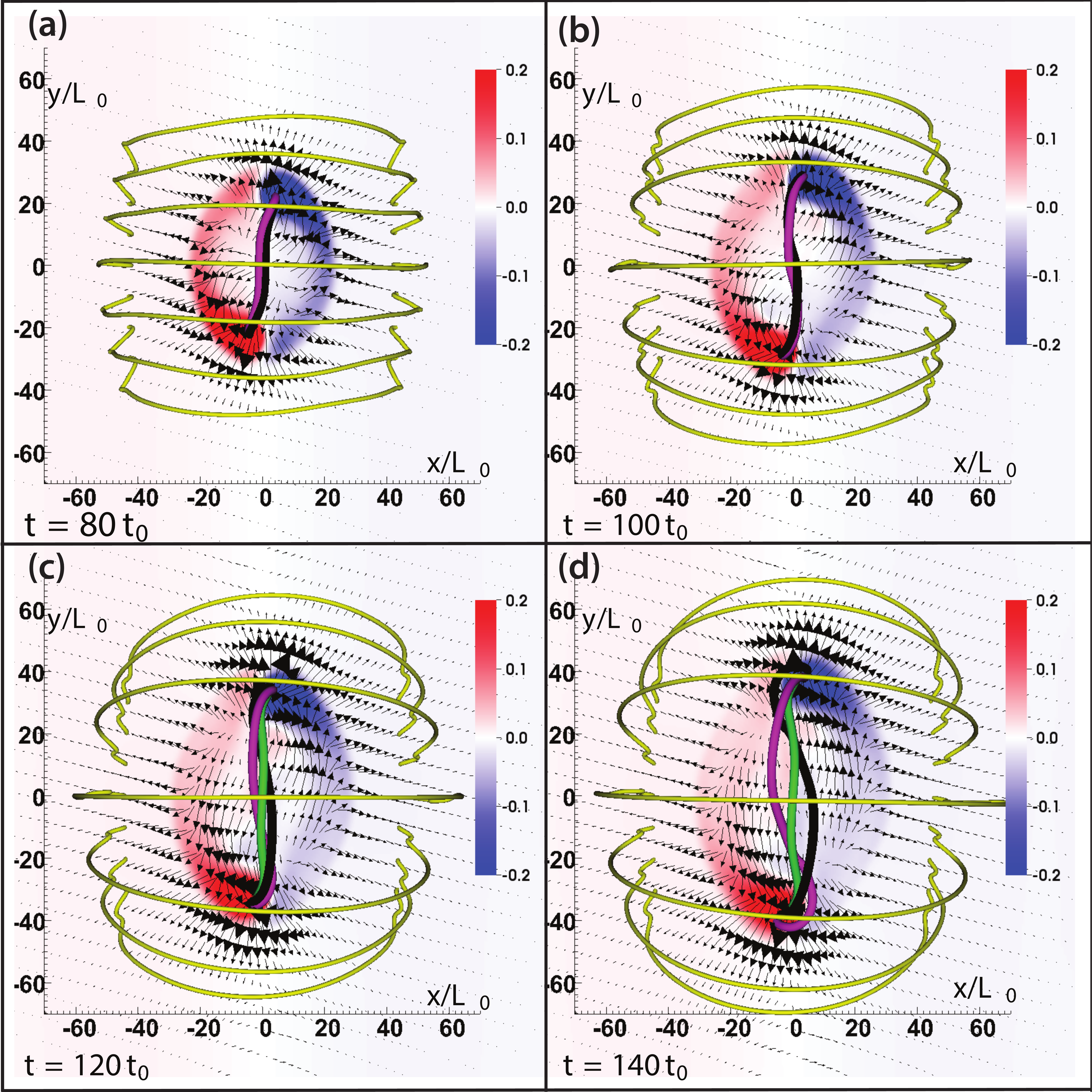}
\vspace{0mm}
\caption{Rotation of sunspots and the formation of a coronal flux rope at times $t=80,100,120$ and $140t_{0}$.  The color shading shows the vertical component of magnetic field in the $z=0$ plane. 
{  The arrows represent horizontal velocities perpendicular to the magnetic field in the $z=0$ plane, and are scaled by magnitude.}
The black and purple fieldlines originate at $\pm \max{y}$, respectively, at the intersection of the original convection zone tube axis and the side boundaries. The yellow lines originate at the lower boundary and belong to the dipole field. The velocity vectors show a strong shearing component, but also suggest a rotational motion near the center of each polarity region. A new flux rope axis can be defined at the location of the O-point in the $y=0$ plane when the black and purple fieldlines separate. This new axis fieldline is shown as the green fieldline.  \label{fig:flux_rope_form_early_1}}
\end{figure*}

A right-hand-twisted magnetic flux tube is inserted at $x=0,z=z_{t}=-12L_{0}$, aligned along the $y$ axis, and is given by:
\begin{eqnarray}
B_{y} & = & B_{t} {e}^{-r^{2}/R^{2}}, \\
B_{\theta} & = & qrB_{y}, \textrm{where} \\
r & = &\sqrt{(x^2+(z-z_{t})^2)}.
\end{eqnarray}
The width of the tube is $R=2.5L_{0}$, and the strength at $r=0$ is $B_{t}=5B_{0}$. The twist parameter is $q=1/R$. Figure \ref{fig:IC_3D} shows some selected magnetic fieldlines from the initial configuration. The superposition of the flux tube and the dipole field is shown in Figure \ref{fig:IC_3D}, Panel (b). {The conventional wisdom of active region formation is that large-scale $\Omega$-shaped flux tubes, which are anchored well below the visible surface, extend through the surface and into the corona. Hence the initial flux tube used here, which is initially horizontal and line-tied at the side boundaries is not a very realistic initial condition. However, by perturbing the density in the initial flux tube in a certain way, an $\Omega$-shaped tube can be created.
The convection zone flux tube is made buoyant at the center $y=0$, and is neutrally buoyant at its ends at the $y$ boundaries. This is done by perturbing the background density $\rho_{0}(z)$ and background specific energy density $\epsilon_{0}(z)$ to 
\bea
\rho(r,z) & = & \rho_{0}(z)\left(1 + \frac{p_{1}(r)}{p_{0}(z)}e^{-{\frac{y}{\lambda}^2}}\right)  \textrm{and} \\
\epsilon(r,z) & = &  \frac{(p_{0}(z) + p_{1}(r))}{ \rho (\gamma -1 )},
\eea
where $\lambda = 10L_{0}$, $p_{0}(z)$ is the original pressure profile, and $p_{1}(r)$ is determined by solving $\nabla p_{1}(r) = \mathbf{j}\times\mathbf{B}(r)$ for the flux tube's field. As will be shown later, this creates sub-photospheric `legs' of the emerging tube which have a significant vertical component. To optimize the confining effect of the dipole field, the direction of the dipole field is chosen so that it is aligned with the fieldlines in the top edge of the flux tube, i.e.,  $B_{x,dip}>0$. }

It is worth making a point here regarding the use of the phrases `flux tube' and `flux rope'. In previous studies of flux emergence, `flux tube' has been used to describe the sub-surface initial magnetic field configuration, and `flux rope' has been used to describe the presence of a collection of fieldlines wrapped around a central fieldline we designate as the axis. 
In that sense the original sub-surface flux tube is also a flux rope. In this paper we adopt the previously accepted practice of calling the original sub-surface field configuration a `flux tube', and the twisted coronal structure that is formed during the emergence process a 'flux rope'.

\begin{figure*}[t]
\centering
\includegraphics[width=0.85\linewidth]{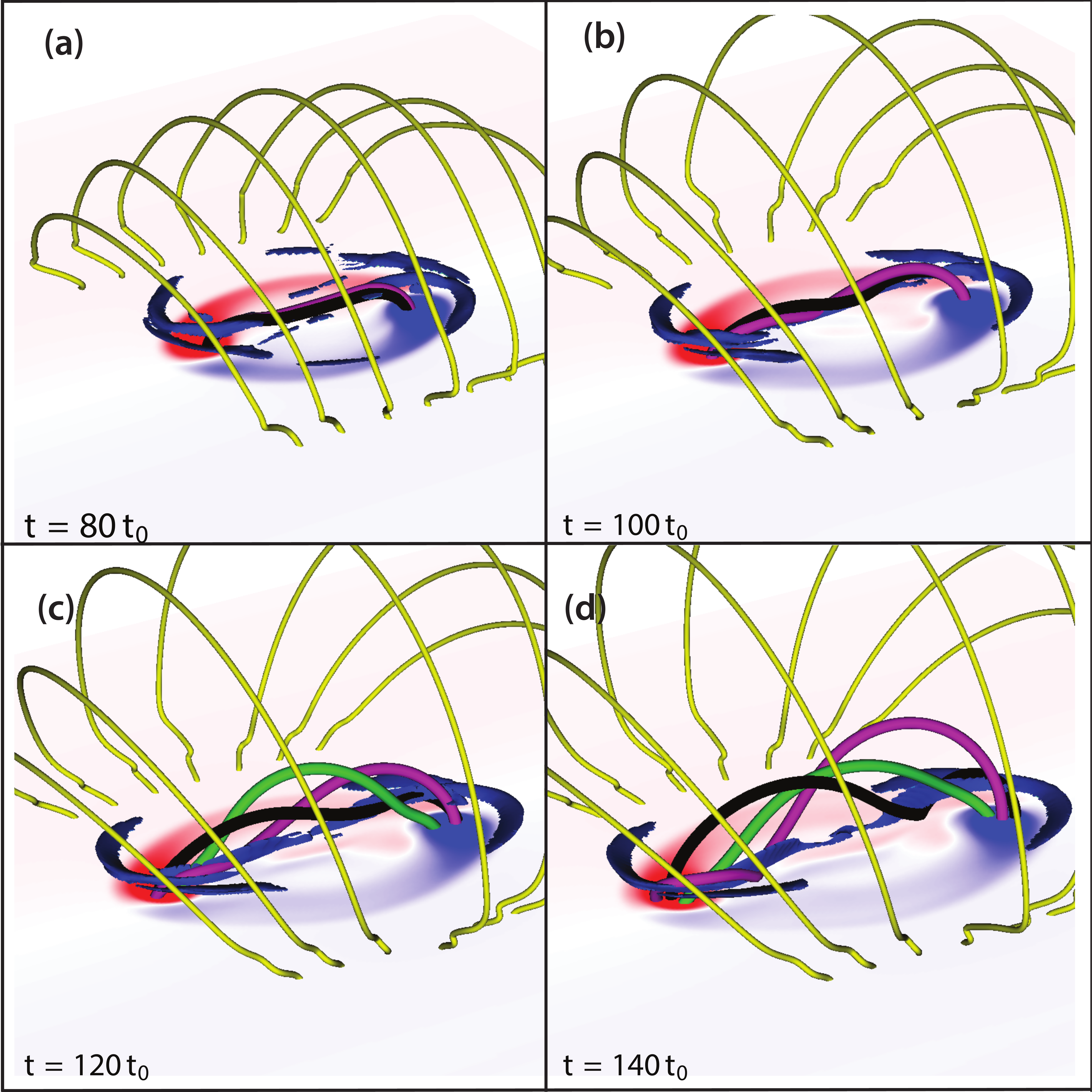}
\vspace{-0mm}
\caption{Same as Figure \ref{fig:flux_rope_form_early_1} but from a different viewpoint. In addition, blue iso-surfaces of  $j \ge 0.035j_{0}$ are plotted for $z \ge 5L_{0}$.
\label{fig:flux_rope_form_early_2}}
\end{figure*}

\section{Results}
\label{Results}
\subsection{Partial Emergence of a Sub-surface Flux Tube}
The partial emergence of a sub-surface flux tube into the solar atmosphere  has been studied and commented upon in a number of previous studies \citep{2001ApJ...554L.111F,2004A&A...426.1047A,2004ApJ...610..588M,2008A&A...479..567M, 2009A&A...507..995M} and we direct the reader to those studies for a more detailed description. The salient points are these: 
The flux tube rises buoyantly until it reaches the convectively stable photosphere/chromosphere, where it temporarily halts and undergoes a large amount of horizontal expansion. Then the upper portions of the deformed tube emerge via the magnetic buoyancy instability \citep{1979SoPh...62...23A} through the photosphere/chromosphere, transition region, and into the corona.

The emergence through the surface of the rising flux tube in Simulation SD is shown in Figure \ref{fig:Early_FE_1}. Note that the boundary conditions employed here allow the same fieldlines to be tracked throughout the simulation to a good approximation by using the same seed point on the side ($y=\pm\max(y)$) boundary. Unless stated otherwise the same fieldlines are drawn for each panel in a given figure. The black line in Figure \ref{fig:Early_FE_1} is the fieldline which  intersects the side ($y=\pm\max{y}$) boundaries at the location of the original convection zone tube axis (note that at this early stage this is one single fieldline). 

\begin{figure}[t]
\centering
\includegraphics[width=1. \linewidth]{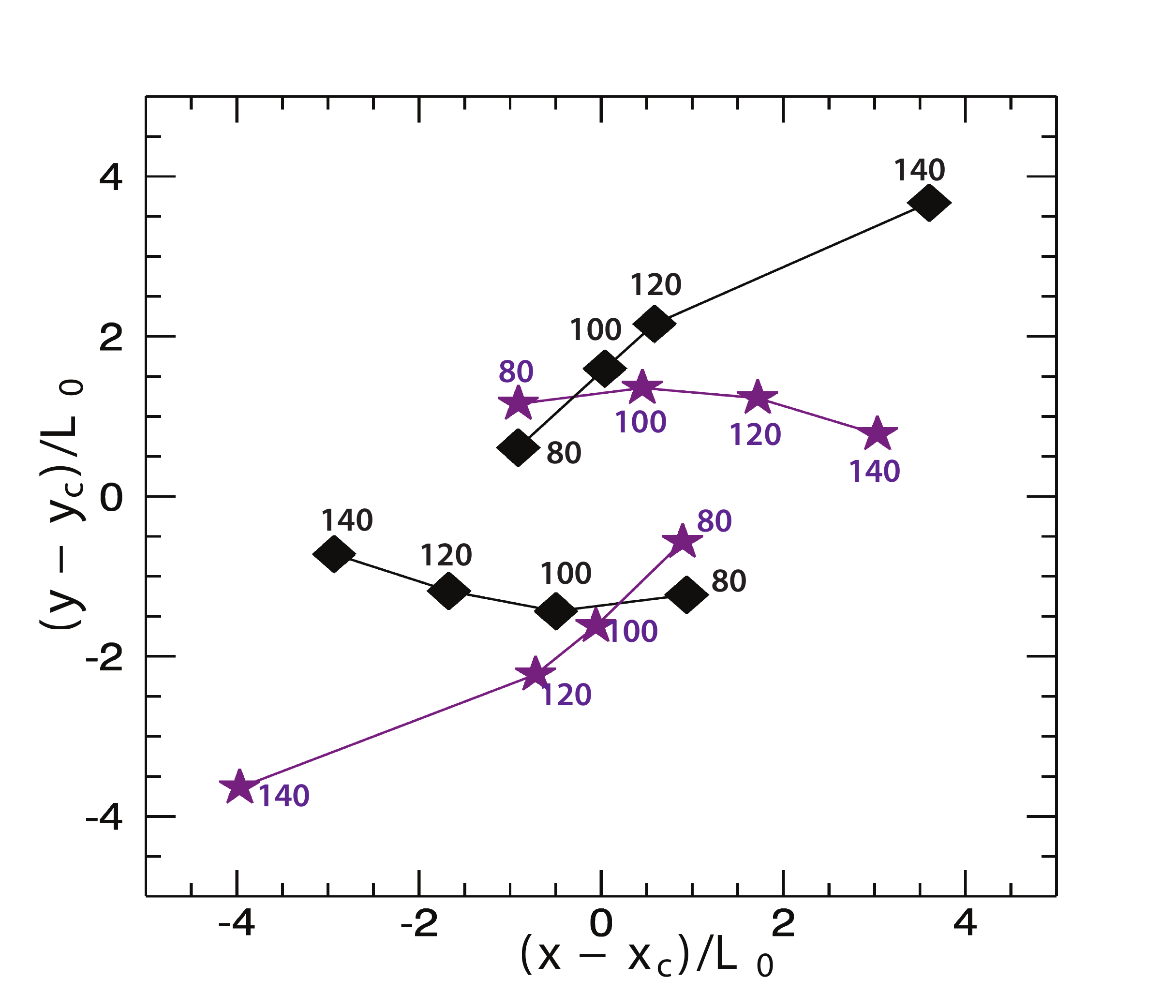}
\vspace{-0mm}
\caption{ {  $(x,y)$ locations  of the intersection of the two axial fieldlines (black and purple fieldlines in Figure \ref{fig:flux_rope_form_early_1}) with the surface ($z=0$), at different times ($t/t_{0}$) in Simulation SD. The black diamonds are for the black fieldline and the purple stars are for the purple fieldline. The locations are taken relative to the center of the polarity region, which is defined as the $(x,y)$ location where the fieldline that goes through the central O-point in the corona (the green line in Figure \ref{fig:flux_rope_form_early_1}) intersects the surface.  The results from both polarity regions are superposed to make one plot. Hence the points in the top half ($y-y_{c} > 0$) are from the intersection of the black and purple fieldlines with the surface in the polarity region in the $y>0$ domain, and the points in the bottom half are from the polarity region in the $y<0$ domain.}
\label{fig:vtheta}}
\end{figure}

\begin{figure*}[t]
\centering
\includegraphics[width=0.85\linewidth]{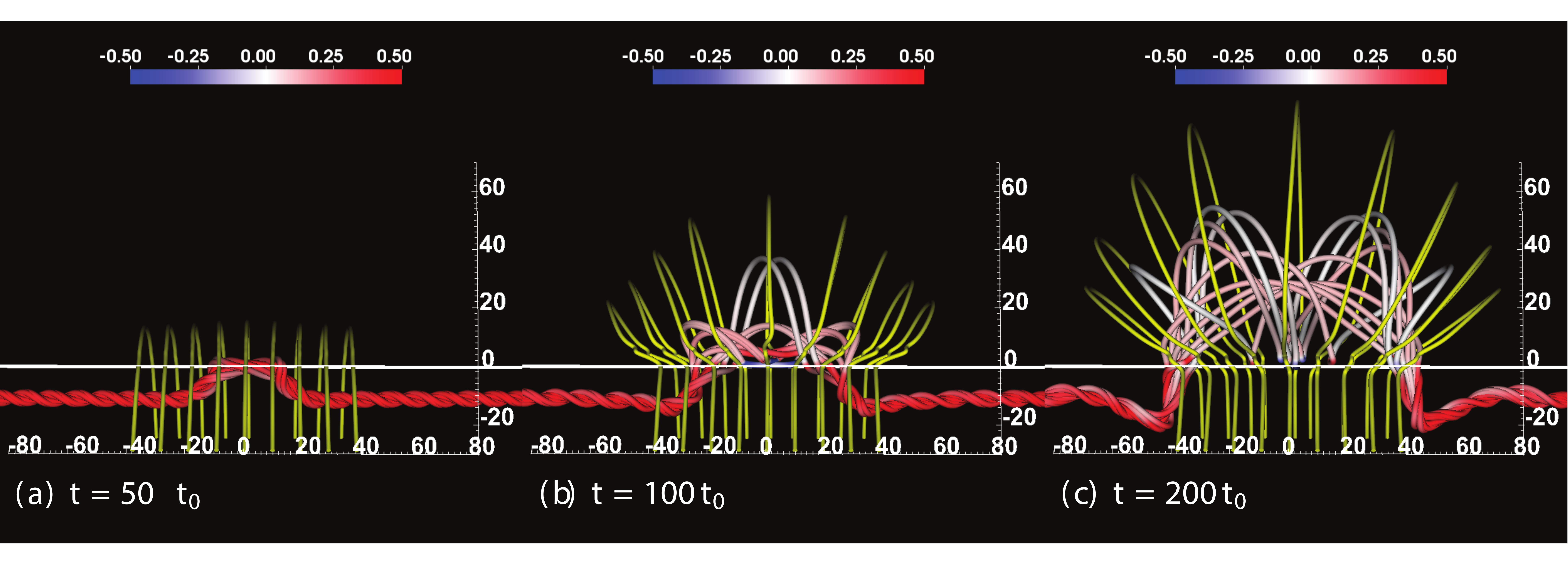}
\vspace{-0mm}
\caption{Twist along fieldlines at times $t=50t_{0}$, $t=100t_{0}$, and $t=200t_{0}$ for simulation SD. Red-white-blue fieldlines are colored with $\alpha L_{0}=\mu_{0}L_{0}\mb{j}\cdot\mb{B}/{|\mb{B}|}^{2}$ and originate from both side ($y=\pm\min{y}$) boundaries. The solid yellow fieldlines originate at the base of the domain and belong to the magnetic dipole field. This figure demonstrates how the emergence of field into the corona causes a gradient in twist along a fieldline as it goes from the convection zone into the corona.  \label{fig:twist_visit}}
\end{figure*}

As shown in Figure \ref{fig:Early_FE_1}, Panel (a), at $t=35t_{0}$ the upper field lines of the flux tube have penetrated the surface ($z=0$) from below. These fieldlines have a high tilt relative to the axis of the tube (the $y$ axis) and they create a  bipolar structure on the surface with a neutral line parallel to the axis of the tube.
 As time progresses, fieldlines emerge with less tilt, i.e., more aligned with the axis of the flux tube. This creates an apparent shearing of the bipole, as shown in Figure \ref{fig:Early_FE_1}, Panels (e) and (f), and the two polarity regions drift apart. This behavior is representative of the observed evolution of emerging active regions \citep{Luoni_2011}.
The emerging field pushes the pre-existing dipole field both vertically and horizontally. These upper fieldlines of the flux tube are nearly parallel to the pre-existing dipole field and so this minimizes the amount of reconnection between the two flux systems.

Figure \ref{fig:Early_FE_2} shows the same fieldlines at later times, but with the color shading of $B_{z}$ at the surface now saturated to highlight the neutral line (which appears white between red and blue). The sections of the fieldlines which cross above the axis of the flux tube (black line)  are concave down. These sections of fieldlines are able to rise further as they drain mass. Beneath the emerged axis fieldline, the sections of the fieldlines are concave-up. These sections carry mass which cannot be drained and are therefore unable to rise further into the atmosphere, as originally found in the simulations of \citet{2001ApJ...554L.111F} and \citet{2004ApJ...610..588M}.
As a result, the  original flux tube emerges only partially:  the sections of the field that are concave down can expand into the corona, while the sections beneath the axis that are concave up remain trapped near the surface. At $t=60t_{0}$, the original flux tube axis has emerged to $3L_{0}$ above the surface, and there is an O-point above the surface in the $y=0$ plane that this axis goes through.
 Previous authors have reported on the location of the original flux tube axis and found that for similar flux tube parameters as those used in this paper, the flux tube axis remains close to or below the surface, typically less than $3L_{0}$ above the surface. \citep{2001ApJ...549..608M,2006A&A...460..909M,2001ApJ...554L.111F}. However, these simulations do not explore the later evolution of the flux tube axis. As we shall show, the original axis of the convection zone flux tube splits into two new fieldlines and these new fieldlines twist around a new coronal flux rope axis.
  
\subsection{Formation of a coronal flux rope}

Figures \ref{fig:flux_rope_form_early_1} and \ref{fig:flux_rope_form_early_2} show the active region for simulation SD at times $[80,100,120, 140]t_{0}$.  The color shading show the vertical component of magnetic field in the $z=0$ plane. {  Figure \ref{fig:flux_rope_form_early_1} also shows the horizontal velocity perpendicular to the magnetic field, which excludes flows caused merely by plasma draining along field lines, on the $z=0$ plane as vectors. }
Also shown in Figures \ref{fig:flux_rope_form_early_1} and \ref{fig:flux_rope_form_early_2} are selected fieldlines. The yellow fieldlines are the dipole field, and originate at the lower boundary.	
The black and purple fieldlines are line-tied at, and originate from, the $y=\pm\max{y}$ boundaries, respectively, at the location of the original flux tube axis. These two fieldlines are coincident early in the simulation, and pass through the O-point located in the $y=0$ plane just above the surface as the original flux tube partially emerges. {  As can be seen in Figures \ref{fig:flux_rope_form_early_1} and \ref{fig:flux_rope_form_early_2} these two fieldlines, which were once the same, separate (perhaps due to magnetic diffusion) and appear to twist about each other as time progresses. They also rise higher into the corona as they do so. }

It should be noted that due to a non-zero value of resistivity, the tracking of fieldlines cannot be exact (even with zero resistivity there is some numerical diffusion to the scheme used to solve the induction equation). We have performed an additional simulation with  ideal $\eta=0$ and found that this splitting of the original axis fieldline also occurs, which suggests it may be independent of the choice of resistivity used.

{  Figure \ref{fig:vtheta} shows the $(x,y)$ locations of the intersection of the two former axial fieldlines with the surface ($z=0$), at different times in Simulation SD. The black diamonds (purple stars) represent locations for the black (purple) fieldline in Figure \ref{fig:flux_rope_form_early_1}. The locations 
are taken relative to the center of the polarity region, which we define as the $(x,y)$ location where the fieldline that goes through the central O-point in the corona (the green line in Figure \ref{fig:flux_rope_form_early_1}) intersects the surface. The results from both active regions are superposed onto one single plot.}
Figure \ref{fig:vtheta} shows  a partial rotational motion of these locations around the center of each polarity region, with the same sign of rotation for each polarity region. {  A rotational motion is also suggested from the vectors in horizontal velocity perpendicular to the magnetic field in Figure \ref{fig:flux_rope_form_early_1}.}
These motions reflect the transport of twist from the convection zone into the corona (see below). Since the green field line in Figure \ref{fig:flux_rope_form_early_1} passes through the O-point without exhibiting significant writhe, it can be considered as a good approximation of the axis of the successively forming coronal flux rope. 

Previous simulations have suggested two different mechanisms for the formation of a coronal flux rope during magnetic flux emergence. \citet{Magara_2006} and \citet{Fan_2009} suggested that rotational motions, brought about by an equilibration of twist along emerging fieldlines, can twist up the coronal sections of fieldlines to create a new flux rope. On the other hand, \citet{2004ApJ...610..588M}, \citet{2008A&A...492L..35A},  and \citet{Archontis_Hood_2012} suggested that the reconnection of emerged sections of sheared fieldlines can create twisted fieldlines, resulting in a flux rope structure in the corona. We now briefly discuss these two mechanisms.

\begin{figure}[t]
\centering
\includegraphics[width=1.\linewidth]{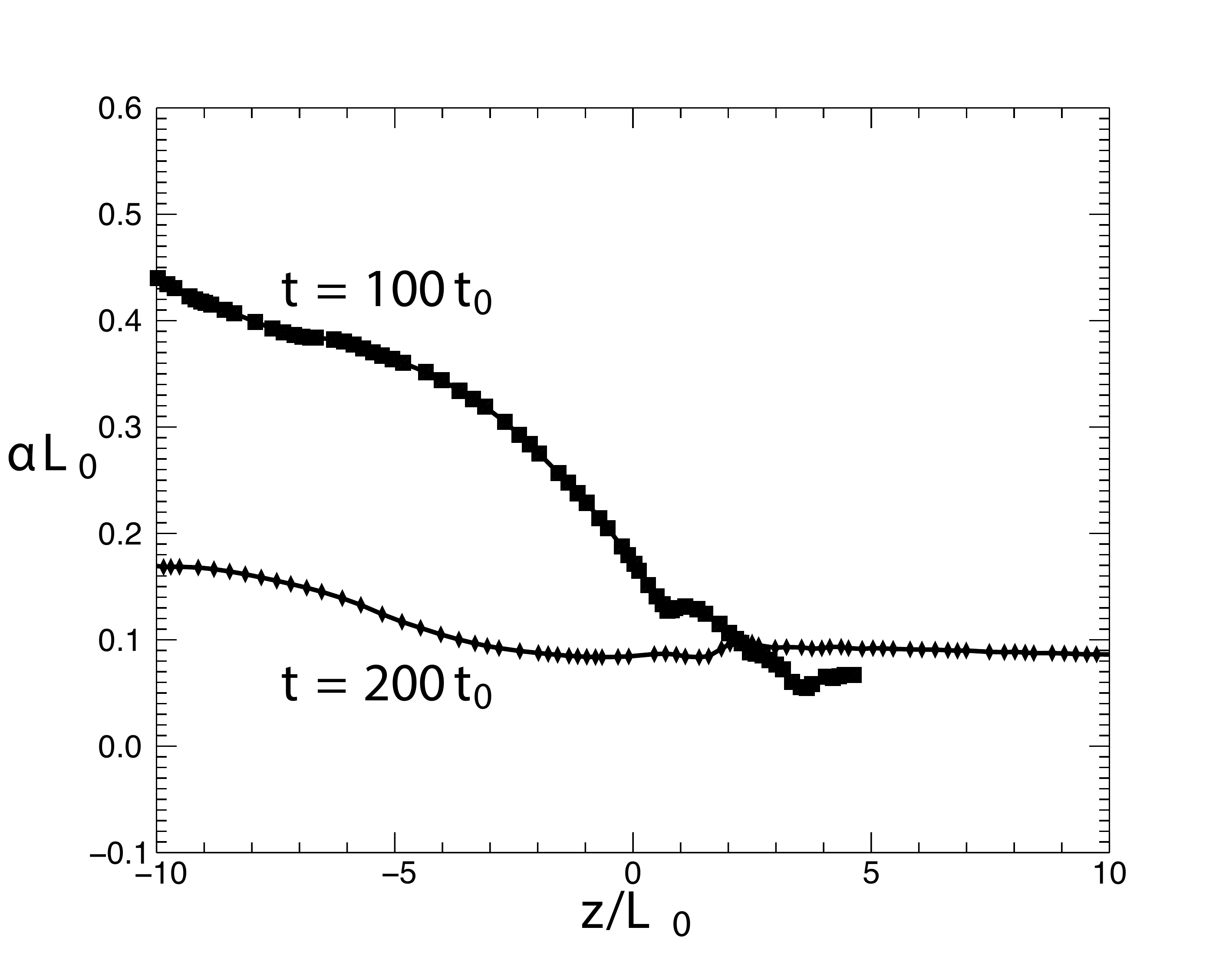}
\vspace{-0mm}
\caption{$\alpha L_{0}$ as a function of $z$ along a fieldline at two different times in the simulation SD, $t=100t_{0}$ and $t=200t_{0}$. The fieldline is the same for both times, and originates at the location of the original convection zone flux tube axis on the $y=\min{y}$ boundary (the purple fieldline in Figures \ref{fig:flux_rope_form_early_1} and \ref{fig:flux_rope_form_early_2}). After $t=80t_{0}$ this fieldline splits from the axis of the flux tube and expands into the corona.  A decrease in the gradient in $\alpha$ between $z=-5L_{0}$ and  $z=0$ can be seen from time $100t_{0}$ to $200t_{0}$. The original value of $\alpha$ along this fieldline at $t=0$ is $0.76/L_{0}$.
 \label{fig:twist_IDL}}
\end{figure}

Figure \ref{fig:twist_visit} shows Simulation SD at times $t=50t_{0}$, $t=100t_{0}$, and $t=200t_{0}$. To give a sense of the local twist per unit length, the fieldlines are colored with the quantity $\alpha L_{0}=\mu_{0}L_{0}\mb{j}\cdot\mb{B}/{|\mb{B}|}^{2}$. As the upper part of the flux tube emerges into the atmosphere, the fieldlines expand into the low-$\beta$ atmosphere, increasing their length. As $\alpha$ is related to twist per unit length and the tube expands faster than the twist propagates upward,  
this creates a gradient in $\alpha$ along the expanding fieldline. Such a gradient was also observed in the simulations of \citet{Fan_2009}, and discussed in \citet{2000ApJ...545.1089L}.
A gradient in twist along a section of a flux tube will drive torsional Alfv\'{e}n waves which equilibrate this twist. Figure \ref{fig:twist_IDL} shows the quantity $\alpha$ at times $t=100t_{0}$ and $t=200t_{0}$, as a function of height, along a portion of the purple fieldline from Figures \ref{fig:flux_rope_form_early_1} and \ref{fig:flux_rope_form_early_2} as it penetrates the surface and passes into the corona. The magnitude of the gradient of $\alpha$ around $z=0$ clearly decreases  in time, indicating that the twist is equilibrating along this section of the flux tube. \citet{Magara_2006} and \citet{Fan_2009} suggested that the torsional motions brought about by this process are capable of causing sunspot rotation which twists up the magnetic field in the corona. This idea is also  supported by recent observations of the formation  of active regions which suggest that sunspot rotation can be attributed to the emergence of twisted magnetic fields \citep{Kumar_2013}.

It has also been suggested that magnetic reconnection is responsible for the formation of coronal flux ropes, by a process similar to what has been suggested based on observations of photospheric flux cancellation \citep{1989ApJ...343..971V}. In flux emergence simulations, the reconnection is driven by a combination of shearing flows, caused by Lorentz forces in the expanding field, and inflows, caused by pressure gradients \citep{2004ApJ...610..588M, 2008A&A...492L..35A, Archontis_Hood_2012}. However, we see no direct evidence of magnetic reconnection, such as an X-point, outflow jets, or curved reconnecting field, underneath the flux rope axis in the simulations described in this paper. This may be due to the combined effects of (i) the limited expansion of the emerging field in the corona due to the presence of the dipole field (which may suppress the amplification of reconnection below the flux rope to a level at which it does not produce noticeable outflow velocities) and (ii) the relatively high resistivity used here, which may suppress the build-up of a steep current layer. We also see no direct evidence of reconnection in the simulation with $\eta=0$, which suggests that the confinement by the dipole field in our simulations, rather than the relatively high resistivity, is the reason that magnetic reconnection beneath the flux rope axis is suppressed. We conclude that the formation process in our simulations is primarily due to the rotation of the polarity regions and the twisting of the field. 

Figure \ref{fig:flux_rope_form_early_2} also shows an iso-surface of current density above $0.03j_{0}$ in the region above $z=5L_{0}$ to highlight the current distribution below the flux rope. At $t=80t_{0}$, the current density is larger above the two regions of concentrated opposite polarity vertical magnetic field
 than above the  center of the bipolar region. After $t=80t_{0}$, there is an increase in current density in the center. The predominant shape of the current sheet when viewed from above is of two distorted J-shapes which merge later to form one S-shape, a process which has been reported in previous flux emergence simulations \citep{Fan_2009, Archontis_Hood_2012}.  Recent extreme ultra-violet observations of  active regions have shown that high temperature (6 MK) J-shaped loops exist before the formation of coronal flux ropes \citep{Liu_2009}, and that these J-structures combine to form a single S-shaped structure when the flux rope is formed \citep[e.g.,][]{2008A&A...481L..65M,2010ApJ...708..314A}. Such S-shaped sigmoid structures have been observed as precursors to CMEs \citep{2000JASTP..62.1427S}.

Figure \ref{fig:flux_rope_form_later_1}  shows the later evolution of Simulation SD. The rotation of the two opposite polarity regions decreases after $t=180t_{0}$, but there is still significant twisting of the fieldlines that extend into the corona. The fieldlines that defined the original convection zone flux tube's axis (black and purple) both wrap around the new flux rope axis in the corona. They also have a pinched U-shape at the center of the active region, which creates the strong current sheet structure.

\subsection{Confinement by overlying field}

\begin{figure*}[t]
\centering
\includegraphics[width=0.85\linewidth]{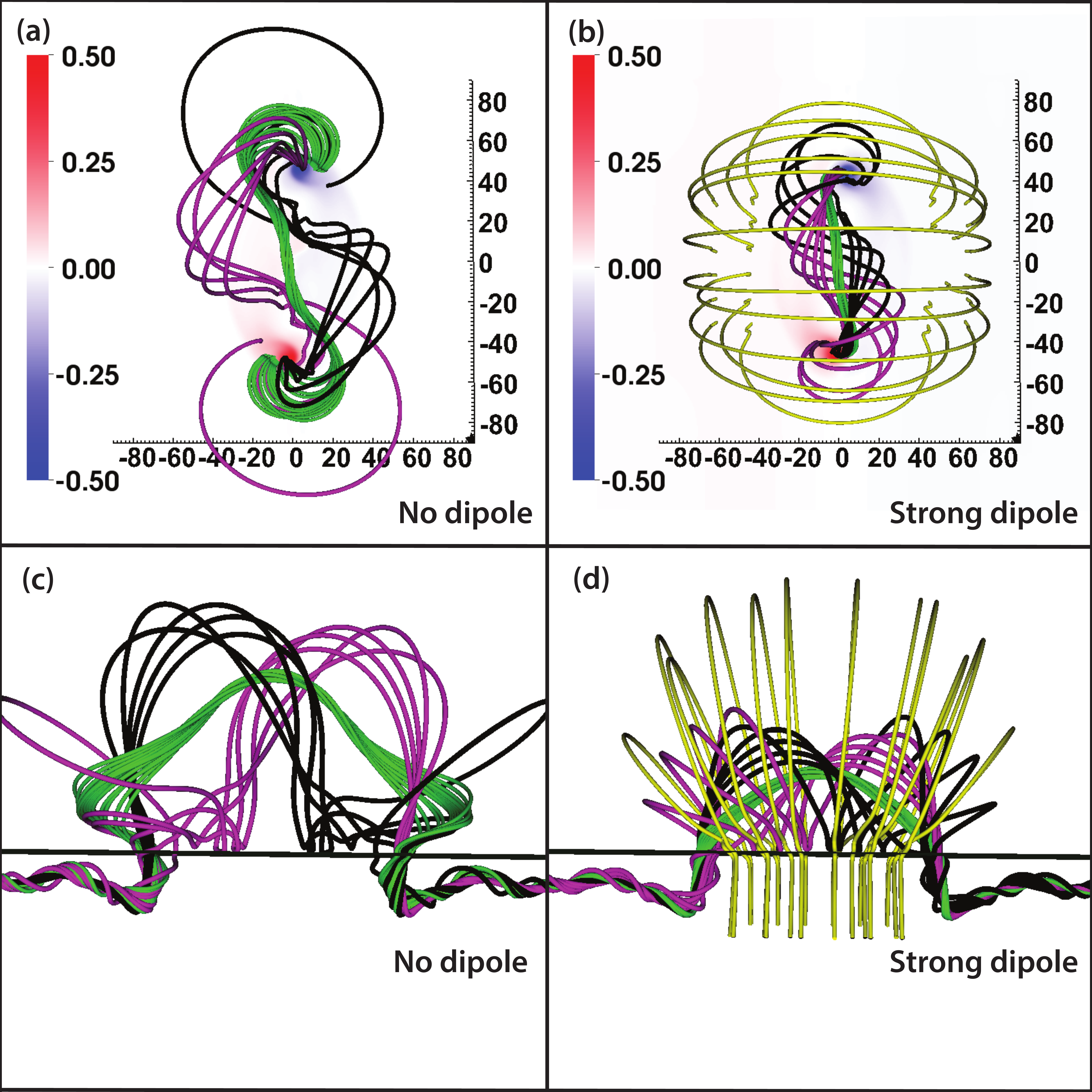}
\vspace{-0mm}
\caption{Comparison of flux rope evolution for simulations without (Simulation ND - Panels a+c) and with (Simulation SD - Panels b+d) an overlying dipole, at time $t=360t_{0}$. The black lines originate from the $y=\max{y}$ boundary around the original convection zone flux tube axis, and the purple lines originate from the same points on the $y=\min{y}$ boundary. The green lines originate close to the new coronal flux rope axis in the $y=0$ plane. The yellow lines in Panels (b) and (d) belong to the dipole field.
 \label{fig:Later_FE}}
\end{figure*}

Figure \ref{fig:Later_FE}  compares the simulation with no dipole (ND) and the simulation with the strongest dipole (SD) at a late stage in the flux rope formation process ($t=180t_{0}$) as the envelope of the flux rope expands further into the corona.  In Simulation SD the dipole field, which was chosen to be aligned so as to minimize reconnection with this envelope field, constrains the expansion (both vertically and horizontally). 

\begin{figure*}[t]
\centering
\includegraphics[width=0.85\linewidth]{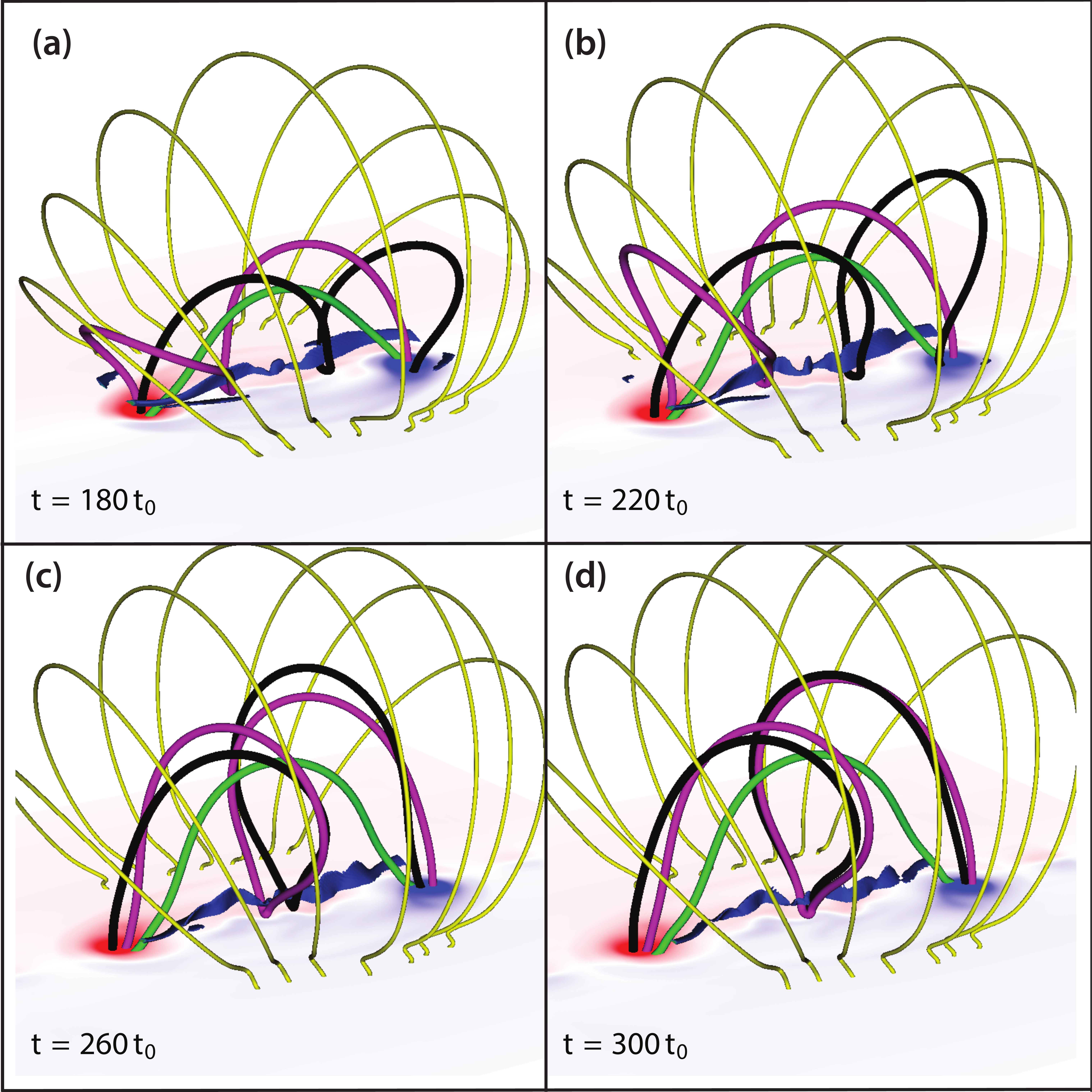}
\vspace{-0mm}
\caption{Same as Figure \ref{fig:flux_rope_form_early_2} but later in time \label{fig:flux_rope_form_later_1}}
\end{figure*}

\begin{figure}[t]
\centering
\includegraphics[width=1.\linewidth]{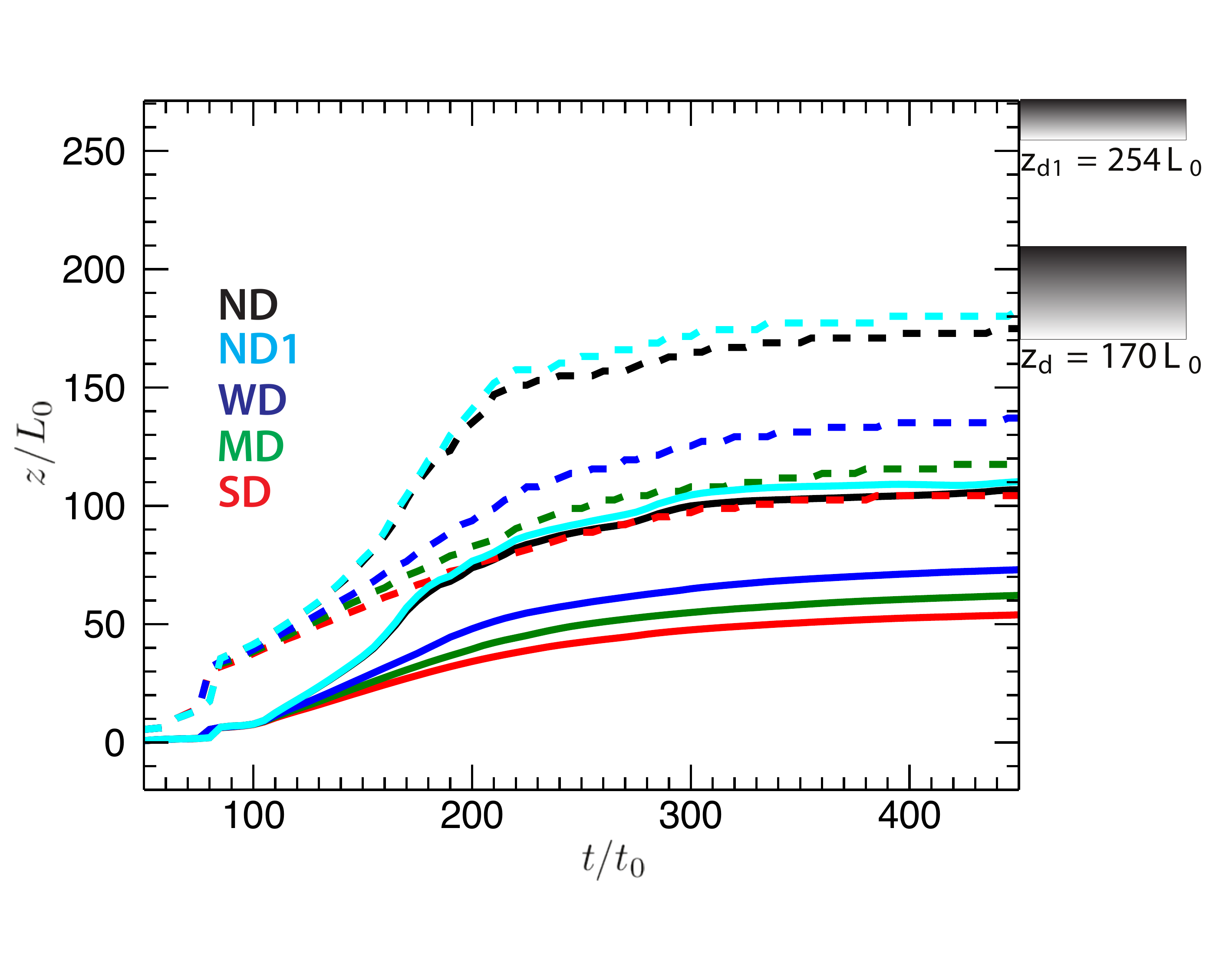}
\vspace{-10mm}
\caption{Height of the O-point in the $y=0$ plane, solid lines, and the height of the envelope field (defined at the height along the $x=y=0$ line where $B_{y}$ is 0.1 times the value at the O-point), dashed lines, for  five simulations (SD, MD, WD, ND and ND1) as a function of time. The two grey shaded boxes represent the vertical extent of the two different damping regions. Simulations SD, MD, WD and ND have a damping region which starts at $z_{d}$ and Simulation ND1 has a damping region which starts at $z_{d1}$ \label{fig:height-time}}
\end{figure}

 Figure \ref{fig:height-time} shows the height of the axis of the coronal flux rope, and the height of the envelope field, as a function of time. The height of the axis of the coronal flux rope is found by locating the point along the $z$ axis at which the horizontal field $B_{x}$ is zero. {  This point is approximately the location of an O-point i.e., where $\sqrt{B_{x}^{2}+B_{z}^{2}}=0$, and the fieldline which goes through this O-point appears to have very little writhe, as shown in Figure \ref{fig:flux_rope_form_later_1}.} The axis of the new flux rope is therefore well represented by this fieldline. We define the height of the envelope field by the intersection of the $z$ axis and the contour of $B_{y}=0.1{B_{y}|}_{axis}$. Ideally we would use the separatrix between the dipole field and the expanding flux rope field to measure the height of the envelope field, but no such separatrix exists in simulation ND. The value of $0.1{B_{y}|}_{axis}$ was chosen because the intersection of  the $z$ axis and this contour is where the separatrix between the dipole field and the expanding flux rope field is located in simulations SD, MD and WD for times $t > 50t_{0}$. 
 
 From $t=200t_{0}$ to $t=450t_{0}$ there is a slow rise of the flux rope, which appears to tend to a stable position. The vertical velocities at the envelope field fall from a typical value of $1.5v_{0}$ at $t=200t_{0}$ to $0.15v_{0}$ at $t=450t_{0}$. As can be seen in Figure \ref{fig:height-time}, the height of the axis of the coronal flux rope at time $t=450t_{0}$ is smaller for larger dipole field strength, as expected.  The simulation with no dipole, Simulation ND, exhibits the strongest expansion, which continues until the envelope field of the flux rope approaches the damping region near  the top boundary. From Figure \ref{fig:height-time} it appears that the flux rope in Simulations SD, MD,  WD, and ND are ultimately stable, but given that the envelope field is so close to the damping region in simulation ND, the effect of the boundary conditions on the stability cannot be ruled out. To investigate this, we perform an additional simulation, ND1, where the top  boundary is extended further out, as is the damping region near this boundary (as described in Section \ref{Num}). We find that envelope field does not extend past $\pm 80L_{0}$ in the $x$ or $y$ directions, and so the the side boundaries, and the damping region at these side boundaries, do not play a role in the stability of the flux rope, Therefore we do not change these in Simulation ND1. Figure \ref{fig:height-time} shows only a small difference in the curves between Simulations ND and ND1. In both cases, the height of the envelope field appears to saturate at $180L_{0}$, which is well below the height of the top damping region for Simulation ND1. We conclude that the confinement of the flux rope in the case of an initially field free corona is not a consequence of the boundary conditions, but of the self-stabilization of the flux rope by its own envelope field.

Previous simulations by \citet{Archontis_Hood_2012} with the same initial tube strength and twist as in the simulations in this paper, and without any pre-existing coronal field, also suggest that the flux rope is ultimately stable. However, the flux rope axis in their simulations reaches a lower height of  $62.3L_{0}$ above the surface compared to the heights of $108L_{0}$ and $110L_{0}$ for the the flux ropes in Simulations ND and ND1 in this paper, respectively. This fact, together with the fact that in the simulations of \citet{Archontis_Hood_2012} the envelope of the flux rope reaches a height of $127L_{0}$ above the surface, at the boundary of the damping region between $127L_{0}$ to $130L_{0}$ above the surface, and thus very close to the top boundary at $130L_{0}$ above the surface, suggests that their boundary conditions are affecting the ultimate height of the flux rope. However, their conclusion, that the coronal flux rope is stabilized by its own envelope field, is supported by our simulations ND and ND1.

In the simulations of \citet{2004ApJ...610..588M}, which used a slightly thinner tube ($w=2L_{0}$ compared to $w=2.5L_{0}$ here), placed initially higher in the convection zone ($z=-10L_{0}$ compared to $z=-12L_{0}$ here), the O-point of the flux rope that formed in the corona rose to a height of $50L_{0}$ by $t=70t_{0}$ at the end of the simulation. This is higher than the flux rope reaches by $t=70t_{0}$ in Simulations ND and ND1, but the flux ropes in Simulations ND and ND1 achieve heights well above $50L_{0}$ later in time, when they become stable.  While the flux rope rises quickly during the initial phase presented in \citet{2004ApJ...610..588M}, based on the findings in this paper, it seems likely that the flux rope formed in \citet{2004ApJ...610..588M} would also ultimately be confined by its own envelope field, if its evolution were followed long enough.

\begin{figure*}[t]
 \centering
\includegraphics[width=0.85\linewidth]{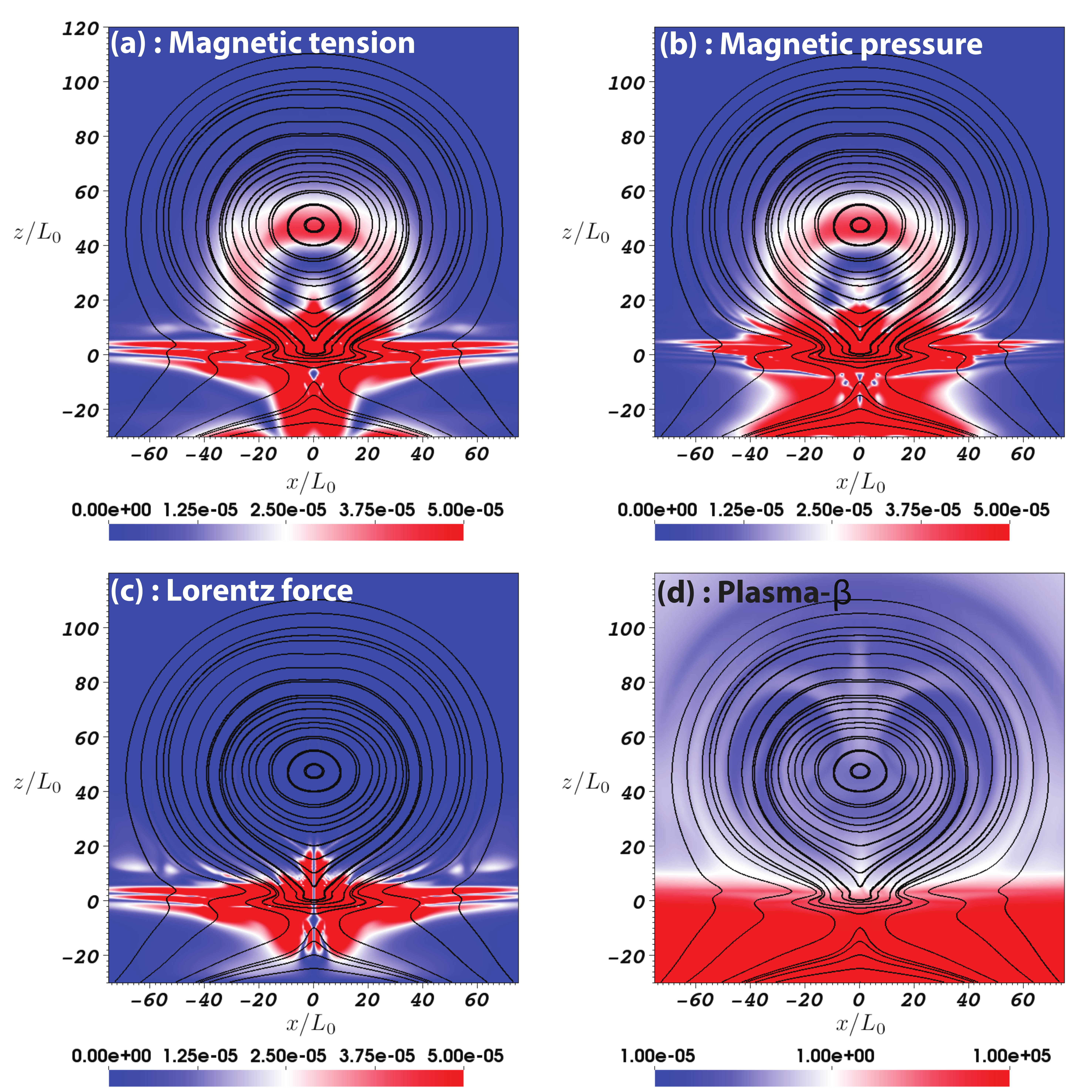}
\vspace{-0mm}
\caption{Magnetic forces and plasma-$\beta$ in the $y=0$ plane for Simulation SD at time $t=280t_{0}$. Panel (a) shows 2D fieldlines in the $y=0$ plane, and a color shading of the magnitude of the 2D magnetic tension vector $(\mb{B}\cdot\nabla B_{x},\mb{B}\cdot\nabla B_{z})L_{0}/B_{0}^{2}$ in the $y=0$ plane. Panel (b) shows the same but for the 2D magnetic pressure vector $(\partial {|\mb{B}|}^{2}/\partial_{x},\partial {|\mb{B}|}^{2}/\partial_{z})L_{0}/B_{0}^{2}$. Panel (c) shows the magnitude of the 2D Lorentz force vector $(
(\mb{j}\times\mb{B})_{x},(\mb{j}\times\mb{B})_{y})\mu_{0}L_{0}/B_{0}^{2}$. Panel (d) shows the plasma-$\beta=P/{|\mb{B}|}^{2}$. This figure demonstrates that in the low-$\beta$ part of the corona, the magnetic tension and magnetic pressure approximately cancel out and so the coronal flux rope is in approximate Lorentz force-balance.  \label{fig:forces}}
\end{figure*}

By including a dipole field in simulations SD, MD and WD, we are also able to constrain the flux rope at lower heights than its own envelope field is able to hold it at. Thus the dipole field is suppressing the rise of the coronal flux rope. The magnetic forces and plasma-$\beta$ in the $y=0$ plane are shown in Figure \ref{fig:forces} for Simulation SD at time $t=280t_{0}$. The magnitudes of the magnetic tension force and the magnetic pressure forces in the $y=0$ plane are shown in Panels (a) and (b), respectively. Above a height of $z=10L_{0}$, the two forces approximately cancel and the magnetic field associated with the flux rope and dipole field has a small Lorentz force relative to the magnitude of the magnetic pressure and tension forces. As can be seen from Panel (d), $z=10L_{0}$ is the height at which the plasma-$\beta$ undergoes the transition from above to below unity. Above $z=10L_{0}$ the magnetic field configuration has approximately $\mb{j}\times\mb{B}=0$.

\subsection{Distribution of Electric Currents}

At present, there is much debate as to whether the electric currents in active regions are `neutralized', in the sense that, for a single sunspot or active region polarity, the direct current (current aligned parallel to the axial magnetic field of the flux tube associated with that sunspot) is surrounded by a return current (aligned anti-parallel to the axial field) which cancels this direct current out. This is important for flare and CME modeling as some models use an initial magnetic field 
with a net current (i.e., return currents are either absent or not large enough to neutralize the direct currents), e.g., the configuration developed by  \citet{1999A&A...351..707T}. Although in the simulations presented here the initial sub-surface flux tube is current-neutralized, i.e., has no net current, there is significant distortion of the magnetic field by the emergence process, and so it is not clear that the resulting coronal flux rope will also be current-neutralized.  

\begin{figure*}[t]
\centering
\includegraphics[width=0.85\linewidth]{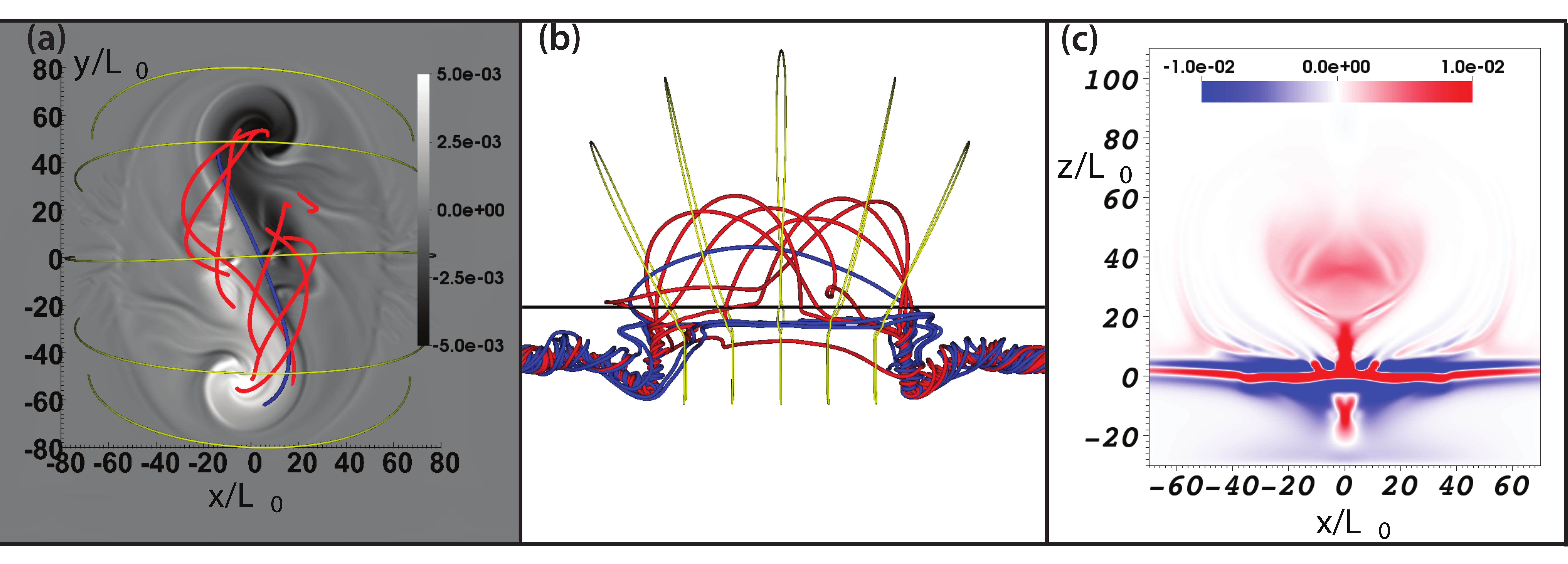}
\vspace{-0mm}
\caption{Distribution of currents in the corona for Simulation SD due to flux emergence and subsequent formation of a coronal flux rope, at time $240t_{0}$. Panels (a) and (b)  show  red current fieldlines originating from regions of direct current ($j_{y}>0$) on the side boundaries, and blue lines originating from regions of return current ($j_{y}<0$), with a gray-scale slice of $j_{z}/j_{0}$ taken at $z=10L_{0}$. The fieldlines are located at regular values of radius with the origin at the axis of the convection zone flux tube, and the number of fieldlines at each radius is proportional to the total axial current at that radius. The yellow lines are the dipole field.  Panel (c) shows $j_{y}/j_{0}$ in the $y=0$ plane.
\label{fig:currents}}
\end{figure*}

To investigate this issue, we plot in Figure \ref{fig:currents} the electric currents at time $t=240t_{0}$ in Simulation SD. Figure \ref{fig:currents}, Panels (a) and (b) show slices of vertical current $j_{z}$ at a height of $z=10L_{0}$, the top of the photosphere/chromosphere region in the model, where  $\beta=1$ for Simulation SD. This height is chosen so as to eliminate any overshoot convective flows that distort the magnetic field. Figure \ref{fig:currents}, Panels (a) and (b) also show current fieldlines (fieldlines of $\mb{j}$). The current fieldlines are colored by the sign of $j_{y}$ at their location of origin on the side boundary: red for direct current ($j_{y}>0$) and blue for return current ($j_{y}<0$). The fieldlines are located at regular values of radius, from $r=0.4L_{0}$ to $6L_{0}$, around the axis of the convection zone flux tube on the side boundary. These radial values are $0.4L_{0} + (0.8nL_{0})$ for $n=0,7$ ($j_{y}$ changes sign at the radial value $2.4L_{0}$). The number of fieldlines at each radial value $r$ is proportional to the total unsigned axial current in the annulus $2\pi rdr$ centered on that radius $r$, where $dr=0.8L_{0}$. The total number of fieldlines is 30, so each fieldline represents 
$1/30$ of the total unsigned axial current in the entire flux rope. 
For a given current fieldline there are only two routes by which it can return to a side boundary. Firstly, it can exit through the opposite boundary. Secondly, it can reverse direction and return to the same boundary in a region of opposite current.  Figure \ref{fig:currents} shows that it is mostly  current fieldlines that
 originate in regions of direct ($j_{y}>0$) current on the side boundary that enter the corona above $z=10L_{0}$. Figure \ref{fig:currents}, Panel (c) shows that a strong central positive $j_{y}$ develops above $z=10$ in the $y=0$ plane. 
{As predicted by the 2.5D model of \citet{2000ApJ...545.1089L} there is a return current which flows along the interface between the sub-photosphere and corona, though the simulations in this paper show that some direct current extends into the corona. The further study of this current layer in both numerical simulations and analytical is required but beyond the scope of this paper.} 
 
Note that a single blue line emerges into the corona in Figure \ref{fig:currents}, Panels (a) and (b). This current fieldline originates from a region of negative $j_{y}$ (but positive $B_{y}$) on the $y=\min{y}$ plane, and so is considered return current. If this fieldline where to follow a simple $\Omega$ shaped path from one boundary to the other, it would intersect the $z=10L_{0}$ plane such that the current normal to that plane $j_{z}$ would be antiparallel to the magnetic field normal to the plane $B_{z}$, and it would be considered return current in the corona. However, this current fieldline, which is representative of many others, performs a complicated circuit, first crossing underneath the active region before passing into the corona and back into the convection zone. This loop-like circuit results in the fieldline having $j_{z}$ parallel to the magnetic field $B_{z}$ on the $z=10L_{0}$ plane. In this sense, for the $z=10L_{0}$ plane, the blue fieldline is a direct current fieldline, even though for its seed location on the $y=\min{y}$ plane it is a return current fieldline. This changing of currents is due to the complicated current structure underneath $z=10L_{0}$, where the currents are far from force-free, and plasma motions can dominate over magnetic forces.

 These results suggest that the coronal flux rope is not neutralized, in the sense that there is not a balance of direct and return current. Of course, there may be very diffuse return currents surrounding this flux rope, and further, more rigorous, analysis is required to determine whether or not the flux rope is indeed un-neutralized. In depth analysis of the neutralization of active region currents in an analogous simulation is presented in \citet{Torok_2013}

\section{Discussion}

The aim of this paper was to use 3D visco-resistive MHD simulations to investigate whether convection zone flux tube emergence could create coronal magnetic field configurations compatible with a flux rope model such as the one developed by \citet{1999A&A...351..707T}, where a net-current coronal flux rope is tethered by overlying potential field.

Consistent with previous simulations, we found that the initial convection zone flux tube partially emerged into the corona; only sections of fieldlines that were able to shed mass were able to emerge. The original flux tube axis first reached a height of 3 Mm above the surface. This is consistent with simulations by \citet{2001ApJ...549..608M}, \citet{2001ApJ...554L.111F}, and \citet{2006A&A...460..909M}. 

As a result of the transport of twist from the convection zone into the corona, 
 torsional motions manifested themselves in co-rotation of the opposite-polarity regions, and effectively twisted up the field in the corona, 
as originally shown by \citet{Fan_2009}. The fieldline associated with the original convection zone flux tube axis separated into two fieldlines due to magnetic diffusion, and became wrapped around a new flux rope axis in the corona. 
 Two distinct J-shaped current layers beneath the new flux rope axis formed, which began to merge during the rotation of the sunspots. This process of emergence and equilibration of twist supports the conclusions from observations that sunspot rotation is driven by twisted flux tube emergence, and that it can cause the formation of sigmoids prior to a solar flare \citep[e.g.,][]{Min_2009,Kumar_2013}. 

No obvious evidence of magnetic reconnection was seen at the location of the current layer below the new coronal flux rope axis, such as the evidence presented in \citet{2004ApJ...610..588M}, \citet{2008A&A...492L..35A}, and \citet{ Archontis_Hood_2012}. In those simulations there was slow, steady reconnection at the location of the current sheet during the expansion of the emerged field in the corona, and this reconnection amplified as the flux rope rose to successively larger heights. Because in the simulations in this paper the emerged field was constrained by the dipole field, we did not see this reconnection stage clearly. Since, however,  we did see evidence of rotational motions in sunspots as suggested by \citet{Fan_2009}, we conclude that the flux rope formation process is predominantly due to these motions in our simulations.

By varying the height of the top boundary and the upper velocity damping region, and finding that the ultimate height of the flux rope axis was unchanged, we removed the effect of the top boundary conditions on the stability of the flux rope, and concluded that even without a dipole field in the corona, the flux rope was constrained by its own envelope field, which support the results by \citet{2008A&A...492L..35A} and \citet{Archontis_Hood_2012}, which were achieved for smaller simulation boxes.

By adding a dipole field, aligned so as to minimize reconnection with emerging field in the corona, we were able to constrain the expansion of the active region into the corona. The stronger the dipole field, the lower the height of the newly formed coronal flux rope, as expected. Such a system of a coronal current-carrying flux rope (or alternatively a strongly sheared arcade) stabilized by an overlying potential field is a canonical configuration believed to produce solar eruptions.
We found that the relatively simple, idealized initial conditions used in our simulations, with a twisted convection zone flux tube emerging into a dipole field representing a decaying active region, is able to robustly produce such a coronal configuration.

A simple analysis of the electric currents suggests that the majority of the return currents did not emerge into the corona, and so a coronal flux rope with a non-neutralized current was created. Further analysis is presented in \citet{Torok_2013}. 
The preliminary results presented here suggest that coronal flux rope models that consider only direct currents, such as the model of \citet{1999A&A...351..707T}, are compatible with the magnetic fields created by the emergence of a twisted magnetic flux tube.

\label{Disc}

\begin{acknowledgments}
\nind{Acknowledgements:}
This work has been supported by the NASA Living With a Star \& Solar and Heliospheric 
Physics programs, the ONR 6.1 Program, and by the NRL-\textit{Hinode} analysis program.
The simulations were performed under a grant of computer time from the DoD HPC program. 
\end{acknowledgments}
\newpage
\bibliographystyle{plainnat}
\bibliography{bibliography}

\end{document}